\documentclass[aps, prl, floatfix, nofootinbib, superscriptaddress, twocolumn]{revtex4-1}

\usepackage{latexsym}
\usepackage{amsmath}
\usepackage{amssymb}
\usepackage{amsfonts}
\usepackage{slashed}

\usepackage[mathscr,scaled=1.15]{urwchancal}
\DeclareFontFamily{OT1}{pzc}{}
\DeclareFontShape{OT1}{pzc}{m}{it}%
{<-> s * [1.15] pzcmi7t}{}
\DeclareMathAlphabet{\mathpzc}{OT1}{pzc}{m}{it}

\usepackage{color}

\usepackage{supertabular}
\usepackage{placeins}
\usepackage{enumitem}
\usepackage{epsfig}
\usepackage{graphicx}

\definecolor{purple}{rgb}{0.5,0,0.5}
\definecolor{blue}{rgb}{0.0,0,0.9}
\definecolor{prdblue}{rgb}{0.133,0.118,0.498}
\usepackage[colorlinks=true, pdfstartview=FitV, linkcolor=prdblue, citecolor= prdblue, urlcolor=prdblue]{hyperref}

\begin{document}

%Symmetry preserving truncations of the gap and Bethe-Salpeter equations
\title{$\,$\\[-7ex]\hspace*{\fill}{\normalsize{\sf\emph{Preprint no}. NJU-INP 029/20}}\\[1ex]
On Mass and Matter}
%Mass at the Core of Matter}
%% 10 pages including references

\author{Craig D.~Roberts}
%\email[]{cdroberts@nju.edu.cn}
\affiliation{School of Physics, Nanjing University, Nanjing, Jiangsu 210093, China}
\affiliation{Institute for Nonperturbative Physics, Nanjing University, Nanjing, Jiangsu 210093, China}

\begin{abstract}
The visible Universe is largely characterised by a single mass-scale; namely, the proton mass, $m_p$.  Contemporary theory suggests that $m_p$ emerges as a consequence of gluon self-interactions, which are a defining characteristic of quantum chromodynamics (QCD), the theory of strong interactions in the Standard Model.  However, the proton is not elementary.  Its mass appears as a corollary of other, more basic emergent phenomena latent in the QCD Lagrangian, \emph{e.g}.\ generation of nuclear-size gluon and quark mass-scales, and a unique effective charge that may describe QCD interactions at all accessible momentum scales.  These remarks are explained herein; and focusing on the distribution amplitudes and functions of $\pi$ and $K$ mesons, promising paths for their empirical verification are elucidated.  Connected therewith, in anticipation that production of $J/\psi$-mesons using $\pi$ and $K$ beams can provide access to the gluon distributions in these pseudo-Nambu-Goldstone modes, predictions for all $\pi$ and $K$ distribution functions are provided at the scale $\zeta=m_{J/\psi}$.
\end{abstract}

%\pacs{ 11.10.St, 11.30.Rd, 12.38.Lg, 24.85.+p }
\date{20 January 2021 $\qquad$ Email: \href{mailto:cdroberts@nju.edu.cn}{cdroberts@nju.edu.cn}}
%\date{12 November 2020}
%\date{26 October 2020}

\maketitle

\noindent\emph{1:\;Introduction}.\,---\,
%%\section{Introduction}
%%\unskip\smallskip
%
In looking at the known Universe, one could be awed by the many complex things it contains.  Even the Earth itself is complicated enough to generate questions in the minds of we observers; basic amongst them are those which focus on our own existence and composition.  Here, too, there are many levels to be explored, right down to the nuclei at the core of every atom and molecule; and even deeper, to the neutrons and protons (nucleons) that constitute those nuclei.  Faced with all this, physicists nevertheless assume that a few succinct mathematical rules should be sufficient to provide a complete explanation of everything we can now perceive and which might become perceptible in future.  That may be correct or it might be hubris \cite{Ellis:2014sjr}; but it would be a bold observer who today offered a definitive answer.

Whether or not Nature can be reduced to an explanation expressed in a few mathematical rules, this approach has been remarkably successful in many areas.  Given that most of the audience will be reading this document on a laptop, after having retrieved it from a remote server, no other illustration is necessary.

So, what is the most fundamental Lagrangian that science has developed to the point that testing is a reality and falsification has thus far been evaded?  Here the definition of ``most fundamental'' might be contentious; but a fair candidate is the Standard Model of particle physics (SM).  This theory was made complete by discovery of the Higgs boson at the large hadron collider in 2012 \cite{Aad:2012tfa, Chatrchyan:2012xdj}, resulting in the subsequent award of the Nobel Prize in Physics to Englert and Higgs \cite{Englert:2014zpa, Higgs:2014aqa} ``\ldots for the theoretical discovery of a mechanism that contributes to our understanding of the origin of mass of subatomic particles \ldots''

As Nature is now understood, the Higgs boson, or something like it,\footnote{It was long ago suggested that all $J=0$ bosons may be \cite{Schwinger:1962tp} ``\ldots secondary dynamical manifestations of strongly coupled primary fermion fields and vector gauge fields \ldots'', in which case the SM's elementary Higgs boson might also be composite.}
is essential to the formation and evolution of our Universe.  For instance, the Higgs mechanism provides large masses to the weak-force gauge bosons, thereby ensuring that weak interactions are short range and protecting against the destabilising influence of electrically charged bosons that propagate over great distances; and making down $(d)$ quarks more massive than up $(u)$ quarks, so helping to ensure stability of the proton and constraining the rate of big bang nucleosynthesis.

Notwithstanding these and many other influences of the Higgs boson, when looking at Nature one is struck by the fact that the vast bulk of visible matter is characterised by a single measurable mass, \emph{viz}.\ $1.673 \times 10^{-27}\,$kg.  This is the proton mass, which translates into natural units (defined such that $\hbar = 1 = c$) as $m_p = 0.939\,$GeV.  Where is the Higgs here?  The masses of the $u$- and $d$-quarks are more than 100-times smaller than $m_p$ \cite{Zyla:2020zbs}; hence, more than 98\% of the proton mass is seemingly ``missing'' from the SM Lagrangian.

The origin of the proton mass, and with it the basic mass-scale for all nuclear physics, is one of the most profound puzzles in Nature.  Since the question addresses the proton, the first place to look is within the strong interaction sector of the SM, \emph{i.e}.\ quantum chromodynamics (QCD), which appeared as the culmination of efforts by a large number of people over many years \cite{Marciano:1979wa, Marciano:1977su}.  %So this is where the next section begins.

\smallskip

\noindent\emph{2:\;Nonperturbative Quantum Chromodynamics}.\,---\,
QCD is a local, Poincar\'e-invariant quantum gauge field theory with interactions built upon the non-Abelian group SU$(3)$.  The Lagrangian is concise:
{\allowdisplaybreaks
\begin{subequations}
\label{QCDdefine}
\begin{align}
{\mathpzc L}_{\rm QCD} & = \sum_{j=u,d,s,\ldots}
\bar{q}_j [\gamma_\mu D_{\mu} + m_j] q_j + \tfrac{1}{4} G^a_{\mu\nu} G^a_{\mu\nu},\\
D_{\mu} & = \partial_\mu + i g \tfrac{1}{2} \lambda^a A^a_\mu\,, \\
\label{gluonSI}
G^a_{\mu\nu} & = \partial_\mu A^a_\nu + \partial_\nu A^a_\mu -
\underline{{g f^{abc}A^b_\mu A^c_\nu}},
\end{align}
\end{subequations}}
\hspace*{-0.6\parindent}where $q_j$ are the quark fields, with $j=u,d,\ldots$ running over the six known quark flavours and $m_j$ being their Higgs-generated current-quark masses; $\{A_\mu^a\,|\,a=1,\ldots,8\}$ are the eight gluon fields, with $\{\lambda^a\}$ being the generators of SU$(3)$ in the fundamental representation; and $g$ is the unique QCD coupling.

It is only the underlined term in Eq.\,\eqref{gluonSI} that fundamenally distinguishes ${\mathpzc L}_{\rm QCD}$ from the Lagrangian of quantum electrodynamics (QED); but whereas QED is ultraviolet incomplete, possessing a Landau pole at some \emph{very large} spacelike momentum (see, \emph{e.g}.\ Ref.\,\cite[Ch.\,13]{IZ80} and Refs.\,\cite{Rakow:1990jv, Reenders:1999bg, Kizilersu:2014ela}), QCD appears empirically to be well-defined at all momenta.  After all, owing to asymptotic freedom \cite{Politzer:2005kc, Wilczek:2005az, Gross:2005kv}, its ultraviolet behaviour is under control; and given our existence, there seem to be no problems at infrared momenta either.  Gluon self interactions, introduced by the underlined term in Eq.\,\eqref{gluonSI}, are certainly the origin of asymptotic freedom and they must also be the source of QCD's infrared stability.

The particular importance of gluon self-interactions was noted long ago.  Consequently, QCD's gauge sector has been the focus of intense scrutiny for more than forty years.  In some quarters, a primary motivation for this attention was the claim that pure-glue QCD possesses a mass gap,\footnote{A prize of \$1\,000\,000 has been offered for a rigorous mathematical proof of the existence of a mass gap in QCD \cite{millennium:2006}.  Notwithstanding the fact that the computer assisted arguments described herein are excluded, they are compelling.}
\emph{i.e}.\ the massless gluons in Eq.\,\eqref{QCDdefine} come to be described by a momentum-dependent  mass-function, whose value at infrared momenta is $m_g = 0.5 \pm 0.2\,$GeV \cite{Cornwall:1981zr}.  If this is correct, then it must be a large part of any answer to the questions surrounding the origin of the proton mass.

The emergence of a gluon mass-scale is surprising for many reasons.  The objection most frequently raised derives from a fear that $m_g\neq 0$ would somehow violate QCD's gauge symmetry.  However, it is readily seen that this is not an issue.  Interaction induced dressing of a gauge boson is expressed in its 2-point Schwinger functions (Euclidean space propagator) through the appearance of a nonzero value for the associated polarisation tensor, $\Pi_{\mu\nu}(q)$.  The generalisation of gauge symmetry to the quantised theory is expressed in Slavnov-Taylor identities \cite{Taylor:1971ff, Slavnov:1972fg}, one of which requires $q_\mu \Pi_{\mu\nu}(q) = 0 = \Pi_{\mu\nu}(q) q_\nu$; but this is ensured so long as
\begin{equation}
\Pi_{\mu\nu}(q) = [\delta_{\mu\nu} - q_\mu q_\nu / q^2] \Pi(q^2).
\end{equation}
In these terms, the interacting gauge boson propagator is expressed as
\begin{equation}
D_{\mu\nu}(q) = [\delta_{\mu\nu} - q_\mu q_\nu / q^2] \frac{1}{q^2[1 + \Pi(q^2)]},
\end{equation}
where any gauge parameter dependence is trivial; hence, omitted here.  A gauge-symmetry-preserving mass-scale appears when $\lim_{q^2\to 0} q^2 \Pi(q^2) =m_g^2$.  This possibility is realised in two-dimensional QED \cite{Schwinger:1962tp} and the effect is now called the Schwinger mechanism of gauge-boson mass generation.

A potentially more powerful objection is found in the observation that if one removes the Higgs-generated current-quark masses from Eq.\,\eqref{QCDdefine}, then the four-dimensional classical action defined by this Lagrangian is scale invariant.  How can a theory that is invariant under arbitrary mass-scale dilations support any unique mass scale?  Here the answer is quantisation \cite{Roberts:2016vyn}.  Local four-dimensional quantum gauge field theories possess ultraviolet divergences.  In order to define any such theory, these divergences must be regularised, following which a renormalisation scheme is introduced to enable the regularisation scale to be traded for renormalised values of couplings and masses.  Physical matrix elements can then be expressed in terms of these renormalised quantities, with true observables being independent of the scheme.

In the renormalisation of a quantum gauge field theory, every quantity that was constant in the classical Lagrangian acquires a dependence on the renormalisation scale, $\zeta$.  Consequently, the trace of the theory's stress-energy tensor, $T_{\mu\mu}$, which is zero in the classical theory, becomes anomalous:
\begin{equation}
\label{SIQCD}
T_{\mu\mu} = \beta(\alpha(\zeta))  \tfrac{1}{4} G^{a}_{\mu\nu}G^{a}_{\mu\nu} =: \Theta_0 \,,
\end{equation}
where $\beta(\alpha(\zeta))$ is QCD's $\beta$-function and $\alpha(\zeta)$ is the associated running-coupling \cite{tarrach}.  Eq.\,\eqref{SIQCD} reveals that a mass-scale exists in every renormalised four-dimensional quantum gauge field theory; but it does not prescribe its size.  The value of $m_g$ in QCD remains a dynamical question.  In principle, $m_g$ could be smaller than the current-masses of the light quarks or larger than the top $(t)$ quark mass; and the answer is not contained within the SM.

\begin{figure}[t]
\centerline{%
\includegraphics[clip, width=0.42\textwidth]{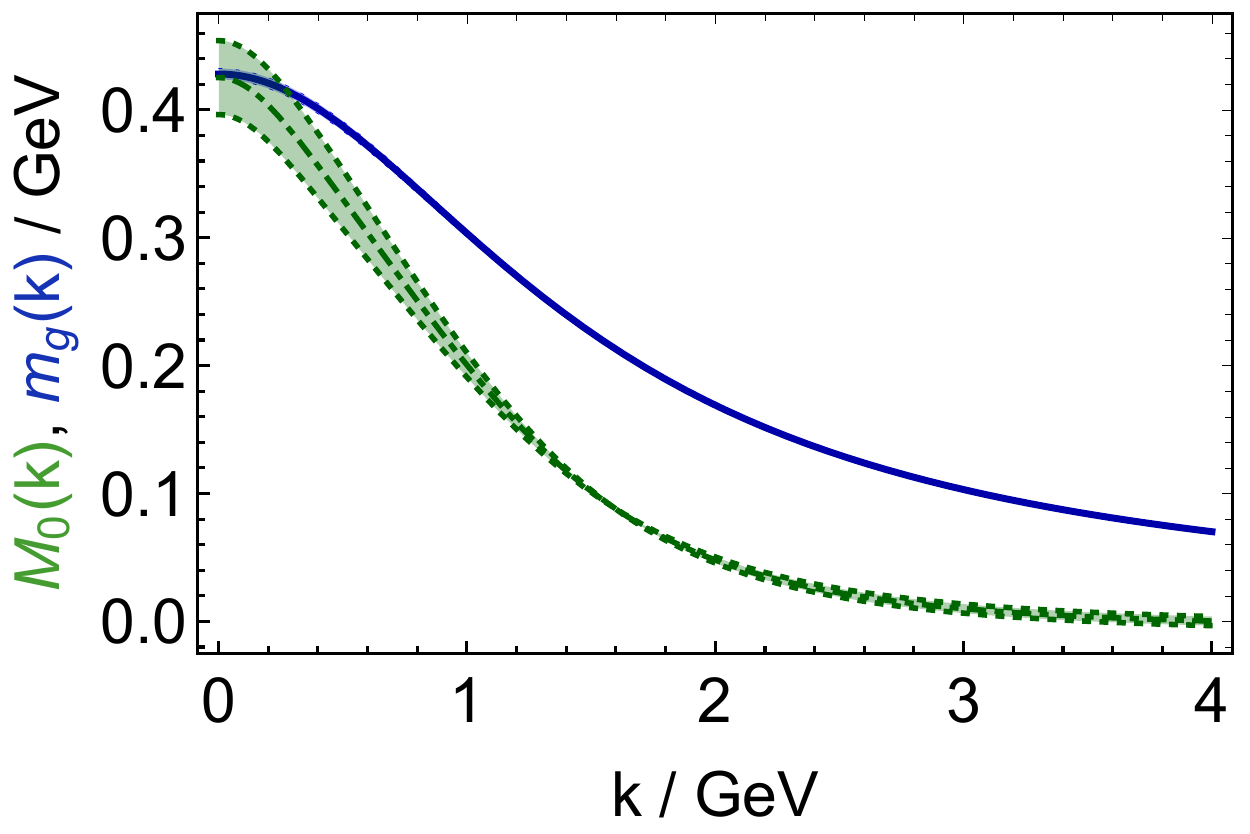}}
\caption{\label{Figmk}
 $m_g(k)$ --  solid blue curve: renormalisation-group-invariant (RGI) gluon running-mass obtained, following the method described in Ref.\,\cite{Aguilar:2019uob}, from the gluon 2-point Schwinger function computed using the lattice-QCD configurations in Refs.\,\cite{Blum:2014tka, Boyle:2015exm, Boyle:2017jwu}.
 %generated with three domain-wall fermions at a physical pion mass and a lattice scale set by the mass of the $\rho$-meson \cite{Blum:2014tka, Boyle:2015exm, Boyle:2017jwu}.
 The barely visible blue band expresses extraction uncertainty from all sources.  (Curve provided by J.~Rodr{\'{\i}}guez-Quintero.)
$M_{0}(k)$ -- dot-dashed green curve: RGI chiral-limit quark running-mass obtained by solving the quark gap equation with the modern kernels described elsewhere \cite{Chang:2011ei, Chang:2013pq, Chang:2013epa}.
%, which were crucial in unifying matter- and gauge-sector focused approaches to understanding QCD \cite{Binosi:2014aea}.
The width of the associated green band expresses existing uncertainties in the dressed gluon-quark vertex \cite{Binosi:2016wcx}.
}
\end{figure}

How can the size of $m_g$ be uncovered?  There is only one answer: methods applicable to nonperturbative phenomena in QCD must be developed to the point that tight links can be drawn between the properties of QCD's gauge sector and measured observables.  This has been the work of forty years so that, today, a combination of tools, exploiting the various strengths of continuum and lattice formulations of QCD, have arrived at a determination of the $\zeta$-independent gluon mass scale \cite{Cui:2019dwv}:\footnote{
This result was obtained using lattice configurations for QCD generated with three domain-wall fermions at a physical pion mass and a lattice scale set by computing the mass of the $\rho$- and $\omega$-mesons \cite{Blum:2014tka, Boyle:2015exm, Boyle:2017jwu}.  It was tested \cite{Zafeiropoulos:2019flq} by verifying that the scale setting choice simultaneously produces a value of the perturbative QCD running coupling at the $Z$-boson mass in agreement with the world average \cite{Zyla:2020zbs}.}
\begin{equation}
\label{gluonmass}
m_0 = 0.43(1)\,{\rm GeV}.
\end{equation}
The associated renormalisation group invariant (RGI, $\zeta$-independent) gluon mass function is depicted in Fig.\,\ref{Figmk}.

There were many important steps along the way to reaching this point, crucial amongst them being a unification of the bottom-up (matter sector based) and top-down (gauge sector focused) approaches to understanding QCD's interactions \cite{Binosi:2014aea}, and perspectives are provided in several contemporary sources \cite{Aguilar:2015bud, Binosi:2016wcx, Gao:2017uox, Huber:2018ned, Fischer:2018sdj, Roberts:2020hiw, Qin:2020rad}.  With both the existence and value of the emergent gluon mass scale having been established by theory, challenges and opportunities arise with the need to elucidate observable consequences: the picture of emergent hadronic mass (EHM) must be tested empirically.

\smallskip

\noindent\emph{3:\;QCD's Running Coupling}.\,---\,
One can first consider the question of a Landau pole in QCD.  At one-loop order in perturbation theory, the QCD running coupling is
\begin{equation}
\alpha(k^2) = \frac{2\pi }{-\beta_1 \ln k^2/\Lambda_{\rm QCD}^2}\,,
\end{equation}
where ``$k$'' is the momentum transfer associated with the process;
$\beta_1 = -C_2^G 11/6 + n_f/3$, with $C_2^G=3$ for SU$(3)$ being a measure of the number of gluon fields ($8=3^2-1$) and $n_f$ counting the active quark flavours;
and $\Lambda_{\rm QCD} \simeq 0.2\,$GeV is the RGI mass scale introduced to align perturbative-QCD predictions with experiment.  (In a complete solution of QCD, the value of $\Lambda_{\rm QCD}$ would be fixed by $m_0$ in Eq.\,\eqref{gluonmass} and vice versa: in the absence of Higgs couplings, the theory has one mass scale, whose value specifies those of all others currently treated as independent.)

The perturbative coupling diverges at $s=\Lambda_{\rm QCD}^2$.  (This is true at all orders in perturbation theory.)  Were that physically true, then measurable cross-sections would exhibit destabilising infrared divergences and we wouldn't be here to observe them.  Hence, if QCD is the correct theory for strong interactions in the SM, then there must be a nonperturbative infrared stabilising mechanism.  The gluon mass scale in Eq.\,\eqref{gluonmass} plays this role.

\begin{figure}[t]
\centerline{%
\includegraphics[clip, width=0.42\textwidth]{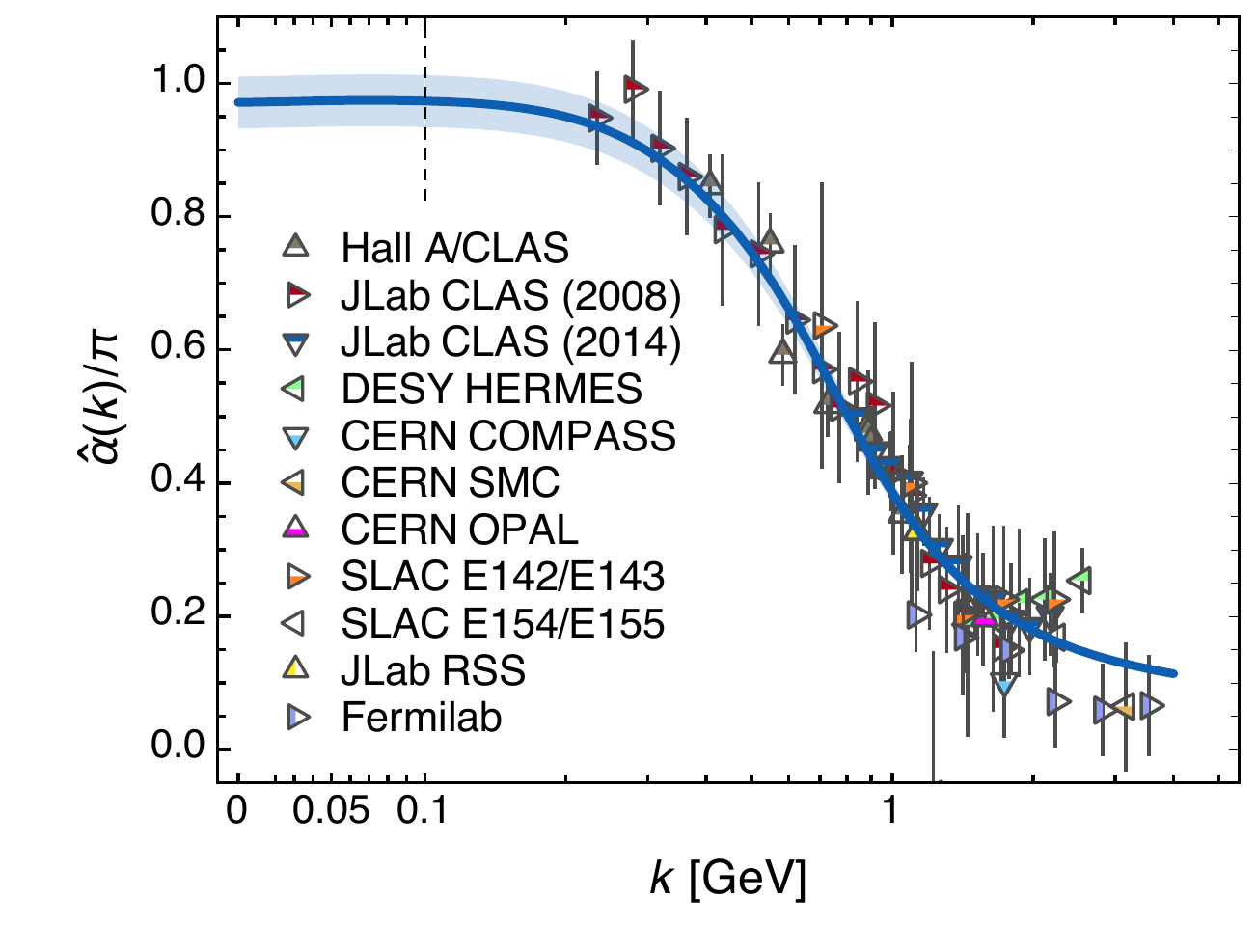}}
\caption{\label{FigalphaPI}
QCD's process-independent running-coupling, $\hat{\alpha}(k^2)/\pi$, obtained by combining the best available results from continuum and lattice analyses \cite{Cui:2019dwv}.
%
%This running coupling saturates at infrared momenta: $\hat\alpha(0)/\pi=0.97(4)$ owing to the dynamical breakdown of scale invariance, expressed through emergence of a gluon mass-scale, with calculated value $m_0/{\rm GeV} = 0.43(1)$.
%
World data on the process-dependent charge $\alpha_{g_1}$ \cite{Deur:2016tte}, defined via the Bjorken sum rule, are also depicted, with sources detailed elsewhere \cite{Cui:2019dwv}.
(Image courtesy of D.\,Binosi.)
}
\end{figure}

Following reconciliation of the bottom-up and top-down approaches to QCD's gauge sector, it became apparent that one could define and calculate a process-independent (PI) running coupling for QCD, $\hat\alpha(s)$ \cite{Binosi:2016nme}.  This charge is a unique QCD analogue of the Gell-Mann--Low effective charge in QED \cite{GellMann:1954fq}, being completely determined by the gluon vacuum polarisation.  The prediction obtained using the most up-to-date continuum and lattice analyses of QCD's gauge sector is depicted in Fig.\,\ref{FigalphaPI}.  Several key features are readily apparent.

\begin{figure*}[!t]
\begin{tabular}{ccc}
\includegraphics[clip, width=0.33\textwidth]{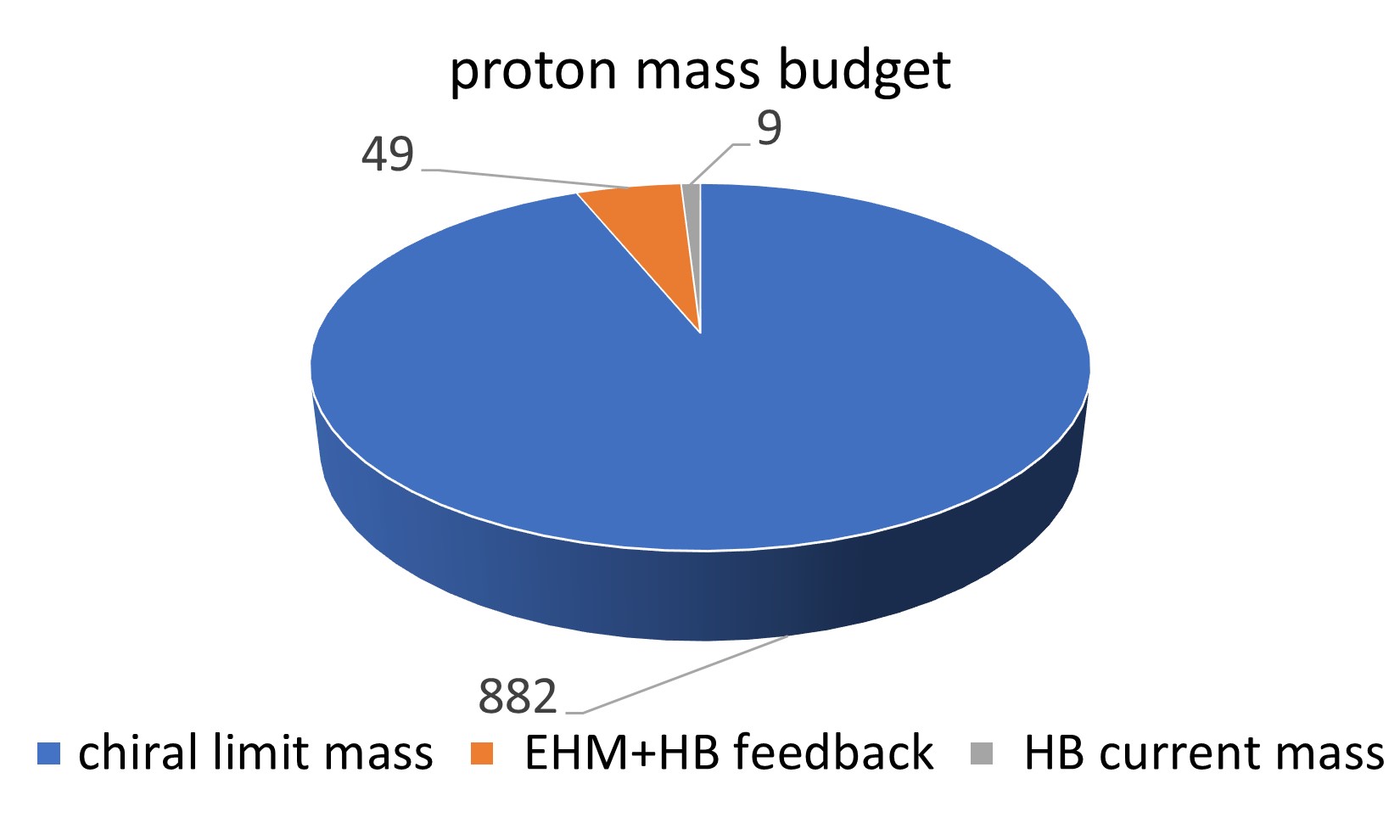} &
\includegraphics[clip, width=0.32\textwidth]{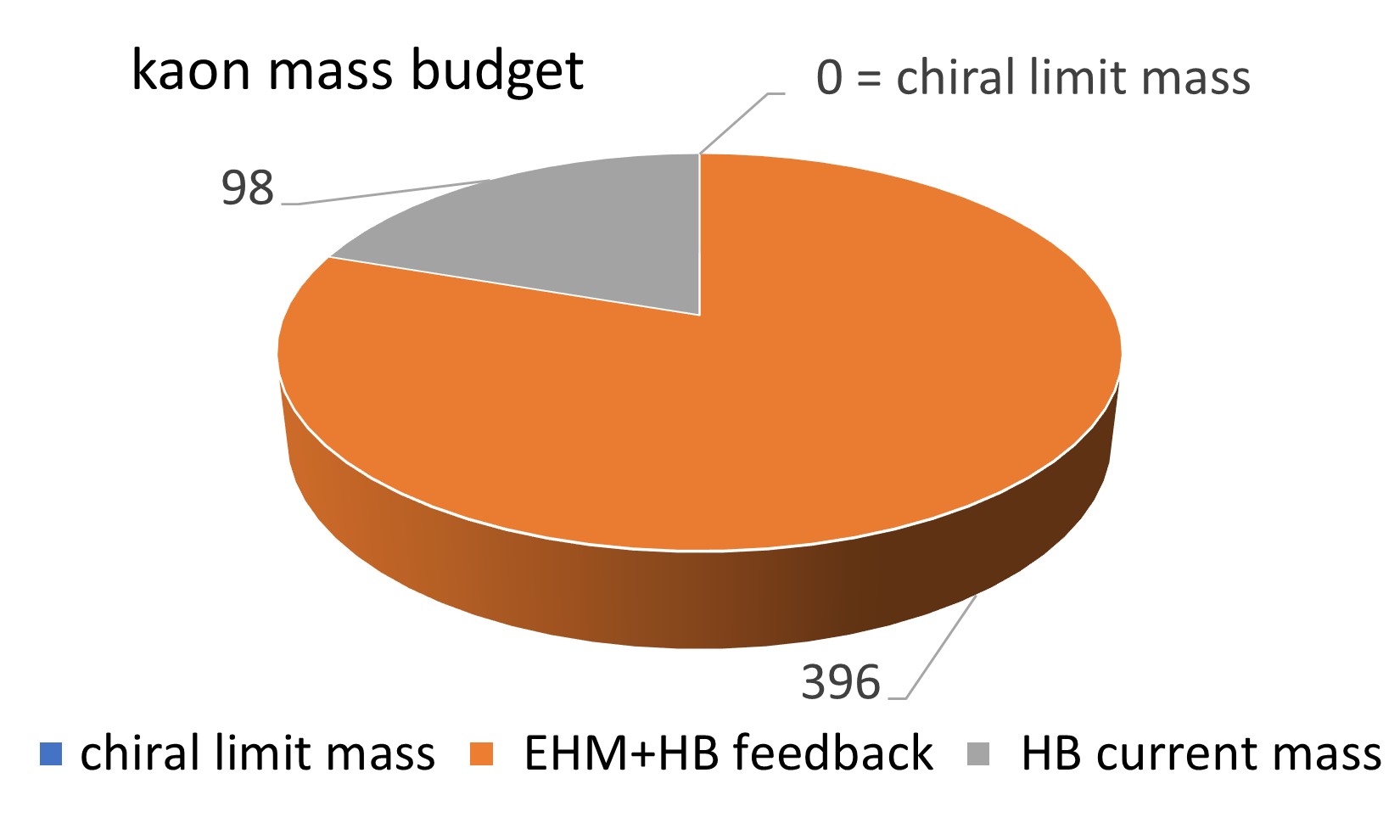} &
\includegraphics[clip, width=0.33\textwidth]{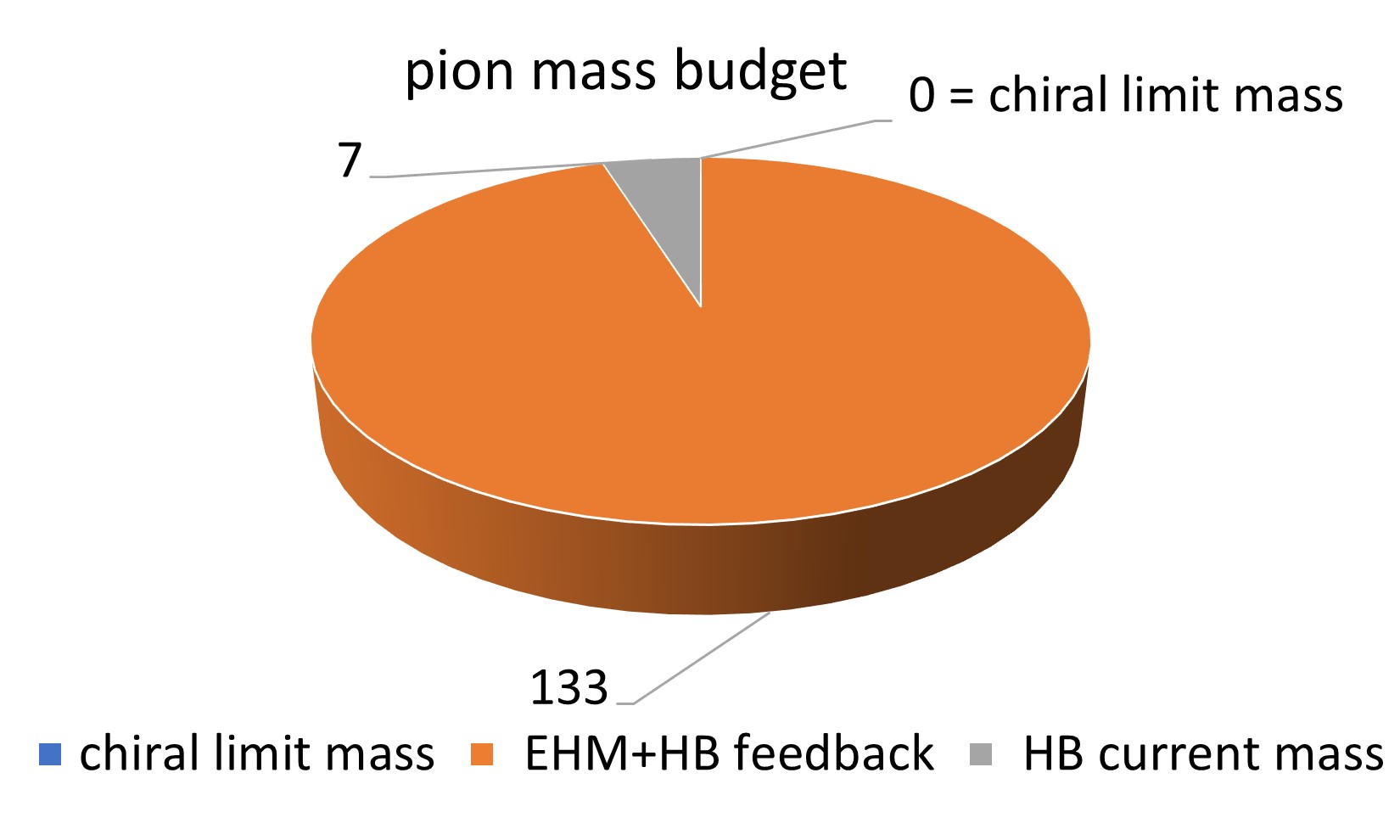}
\end{tabular}
\caption{\label{F1CDR}
Mass budgets for the proton, $K$-meson and $\pi$-meson.  The differences are stark.
Owing to EHM, the proton's mass is large in the chiral limit, as indicated by the blue domain, which constitutes 94\% of $m_p$.
Conversely and yet still owing to EHM via its DCSB corollary, the $K$ and $\pi$ are massless in the absence of quark couplings to the Higgs boson; hence, no blue domain.
Switching on Higgs boson couplings to lighter quarks, two new contributions appear: grey shows the sum of Higgs-generated valence-quark/antiquark current-masses in each hadron; and orange indicates the contribution generated by constructive interference between EHM and Higgs-boson (HB) effects.
5\% of $m_p$ owes to EHM+HB.
On the other hand, EHM+HB interference is responsible for 95\% of the physical $\pi$ mass.
The $K$  lies between these extremes.  It is a would-be NG mode, so there is no blue-domain; but the sum of valence-quark and valence-antiquark current-masses in the $K$ amounts to 20\% of its measured mass -- four times more than in the pion, with EHM+HB interference producing 80\%.
(Units MeV, Poincar\'e-invariant separation at $\zeta=\zeta_2=2\,$GeV, breakdowns produced using information from Refs.\,\cite{RuizdeElvira:2017stg, Zyla:2020zbs}.)
}
\end{figure*}

\begin{enumerate}[label=(\roman*)]
\item The PI coupling is a smooth function of spacelike momenta, saturating in the infrared: $\hat\alpha(s=0)/\pi = 0.97(4)$.  The value of the PI charge at $s=\Lambda_{\rm QCD}^2$, the location of the once-was Landau pole, defines a screening mass: $\zeta_H \approx 1.4 \Lambda_{\rm QCD}$.  On $s\lesssim \zeta_H^2$, interactions between coloured objects are roughly scale invariant; hence, the theory is effectively conformal once again.  These properties owe to the emergence of the gluon mass scale in Eq.\,\eqref{gluonmass}, which ensures that long wavelength gluon modes are screened, playing effectively no dynamical role.  $\zeta_H$ marks a border between soft (nonperturbative) and hard (perturbative) physics.  Hence, it is a natural choice for the ``hadronic scale'', \emph{viz}.\ the renormalisation scale at which one formulates and solves the continuum bound state problem in terms of quasiparticle degrees-of-freedom \cite{Ding:2019qlr, Ding:2019lwe, Cui:2019dwv, Cui:2020dlm, Cui:2020tdf}.

\item So far as available data can reveal, $\hat\alpha(s)$ is practically identical to $\alpha_{g_1}(s)$, the process-\emph{dependent} effective charge \cite{Grunberg:1982fw, Dokshitzer:1998nz, Deur:2016tte} defined in terms of the Bjorken sum rule.  There are sound mathematical reasons for this, explained elsewhere \cite{Binosi:2016nme}.  The Bjorken sum rule provides a basic constraint on knowledge of nucleon spin structure as measured in deep inelastic scattering.  Thus, the link between $\hat\alpha(s)$ and $\alpha_{g_1}(s)$ points to a potentially important role for $\hat\alpha(s)$ in connecting data with calculations of hadron light-front distribution amplitudes and functions \cite{Ding:2019qlr, Ding:2019lwe, Cui:2019dwv, Cui:2020dlm, Cui:2020tdf}.

\item In being process independent, $\hat\alpha(s)$ fulfills a wide range of purposes, unifying a large array of observables.  It is thus a strong candidate for that function which represents QCD's interaction strength at any accessible momentum scale \cite{Dokshitzer:1998nz}.  Moreover, its properties justify a conclusion that QCD is a well-defined quantum field theory in four dimensions.  As such, QCD becomes a candidate for use in SM extensions based on attributing compositeness to particles that may today seem elementary.
\end{enumerate}

\smallskip

\noindent\emph{4:\;Emergence and Evolution of Constituent Quarks}.\,---\,
More than fifty years ago, the constituent quark model (CQM) \cite{GellMann:1964nj, Zweig:1981pd} brought order to a rapidly expanding collection of strong interaction bound states (hadrons): $\pi$, $K$, $\rho$ \ldots mesons; and neutron (n), proton (p), $\Delta$ \ldots baryons.  The approach established that many gross features of the hadron spectrum can be understood by positing the existence of constituent-quarks with nuclear-size masses: $M_U \approx M_D \approx 0.4\,$GeV, $M_S \approx 0.5\,$GeV, etc.  Given the success of this idea, it is natural to ask whether it has a foundation in QCD.  In the past vicennium, an affirmative answer has emerged.

The current-quarks in Eq.\,\eqref{QCDdefine} are strongly interacting.  Thus, compared with free-fermion behaviour, one may expect material changes in their propagation characteristics.  Attempts to compute these changes began with the formulation of QCD \cite{Lane:1974he, Politzer:1976tv}.  They progressively became more sophisticated as experience grew with formulating and solving the quark gap (Dyson \cite{Dyson:1949ha}) equation and as computational methods and power improved for lattice-regularised QCD.  It is now known that even in the absence of Higgs couplings into QCD, quarks acquire a running mass that is large at infrared momenta,  (See, e.g.\, Refs.\,\cite{Bhagwat:2003vw, Bowman:2005vx, Bhagwat:2006tu} and citations thereof.)  This is dynamical chiral symmetry breaking (DCSB), a corollary of EHM: perturbatively massless quarks acquire a large infrared mass through interactions with their own gluon field.

Typical solutions of the quark gap equation are depicted in Fig.\,\ref{Figmk}.  The curves were obtained using modern kernels \cite{Chang:2011ei, Chang:2013pq, Chang:2013epa}, whose development was crucial to arriving at an understanding of QCD's gauge sector \cite{Binosi:2014aea}.  Some quantitative uncertainty remains and is being eliminated as more is learnt about the dressed gluon-quark vertex \cite{Binosi:2016wcx, Bermudez:2017bpx, Aguilar:2018epe, Oliveira:2020yac}; but the gross features are robust:
$M_0(0) \sim m_p/3$; and $M_0(k)$ runs as a logarithm-corrected $1/k^2$ power-law into the ultraviolet.  When Higgs couplings are reinstated, the mass function becomes flavour dependent and its value at the origin is roughly the sum of $M_0(0)$ and the appropriate current-quark mass \cite{Ivanov:1998ms}.

Explaining many of the spectroscopic successes of the constituent-quark picture is now straightforward.  Hadron masses are global, volume-integrated properties.  Hence, studied as bound states in quantum field theory, their values are largely determined by the infrared size of the mass function of the hadron's defining valence quarks \cite{Qin:2019hgk}.  This feature is emphasised by the fact that even a sensibly formulated momentum-independent interaction produces a good overall description of hadron spectra \cite{Yin:2019bxe, Gutierrez-Guerrero:2019uwa}.  The infrared scales needed are provided by the mass function in Fig.\,\ref{Figmk} and related forms for the different quark flavours; and those scales are generated by the effective charge in Fig.\,\ref{FigalphaPI} augmented by Higgs-boson contributions.

Whilst hadron masses are largely insensitive to the running of the dressed-quark mass, this feature becomes vital for dynamical, structural properties, \emph{inter alia}: elastic and transition form factors \cite{Brodsky:2020vco, Carman:2020qmb, Barabanov:2020jvn} and parton distribution functions and amplitudes \cite{Ding:2018xwy, Ding:2019qlr, Ding:2019lwe, Cui:2019dwv, Cui:2020dlm, Cui:2020tdf, Lan:2019rba, Chang:2020kjj, Kock:2020frx}.

\smallskip

\noindent\emph{5:\;Nambu-Goldstone Modes}.\,---\,
There is a class of bound-states whose masses and properties cannot be explained using CQMs; namely, the SM's pseudoscalar mesons: $\pi$, $K$, $\eta$, $\eta^\prime$.  In the absence of Higgs couplings into QCD, the $\pi$ and $K$ mesons are Nambu-Goldstone (NG) modes.  The $\eta$ and $\eta^\prime$ would also be NG modes if it were not for the non-Abelian anomaly \cite{Christos:1984tu}.
In NG modes, the mass-scale that characterises all visible matter is hidden; and its manifestation in the physical $\pi$ and $K$ mesons is very different from that in all other hadrons.  (See, \emph{e.g}.\ Fig.\,\ref{F1CDR} and Ref.\,\cite[Sec.\,V]{Roberts:2020udq}.)  These are two quite particular consequences of EHM: the chiral-limit masking owes to the axial-vector Ward-Green-Takahashi identity and, in the presence of Higgs-quark couplings, the actual meson masses result from constructive interference between EHM and Higgs-boson effects.  Expressed at the most fundamental level within the SM, a necessary and sufficient condition for the existence of NG modes is \cite{Maris:1997hd, Qin:2014vya}
\begin{equation}
\label{EqGTRE}
f_{\rm NG}^0 E_{\rm NG}^0(k,P;P^2=0) = B_0(k^2)\,,
\end{equation}
where, with all quantities evaluated in the chiral limit:
$f_{\rm NG}^0$ is a measure of the NG mode's wave function at the origin in configuration space;
$E_{\rm NG}^0(k,P;P^2=0)$ is the dominant term in the NG mode's Bethe-Salpeter amplitude (a relativistic relative of the Schr\"odinger wave function), with $k$ the relative momentum between the two valence constituents and $P$ the total bound-state momentum;
and $B_0(k^2)$ is the scalar piece of the dressed-quark's self energy, which is simply connected to $M_0(k)$ in Fig.\,\ref{Figmk}.

\begin{figure}[t]
\centerline{\includegraphics[clip, width=0.47\textwidth]{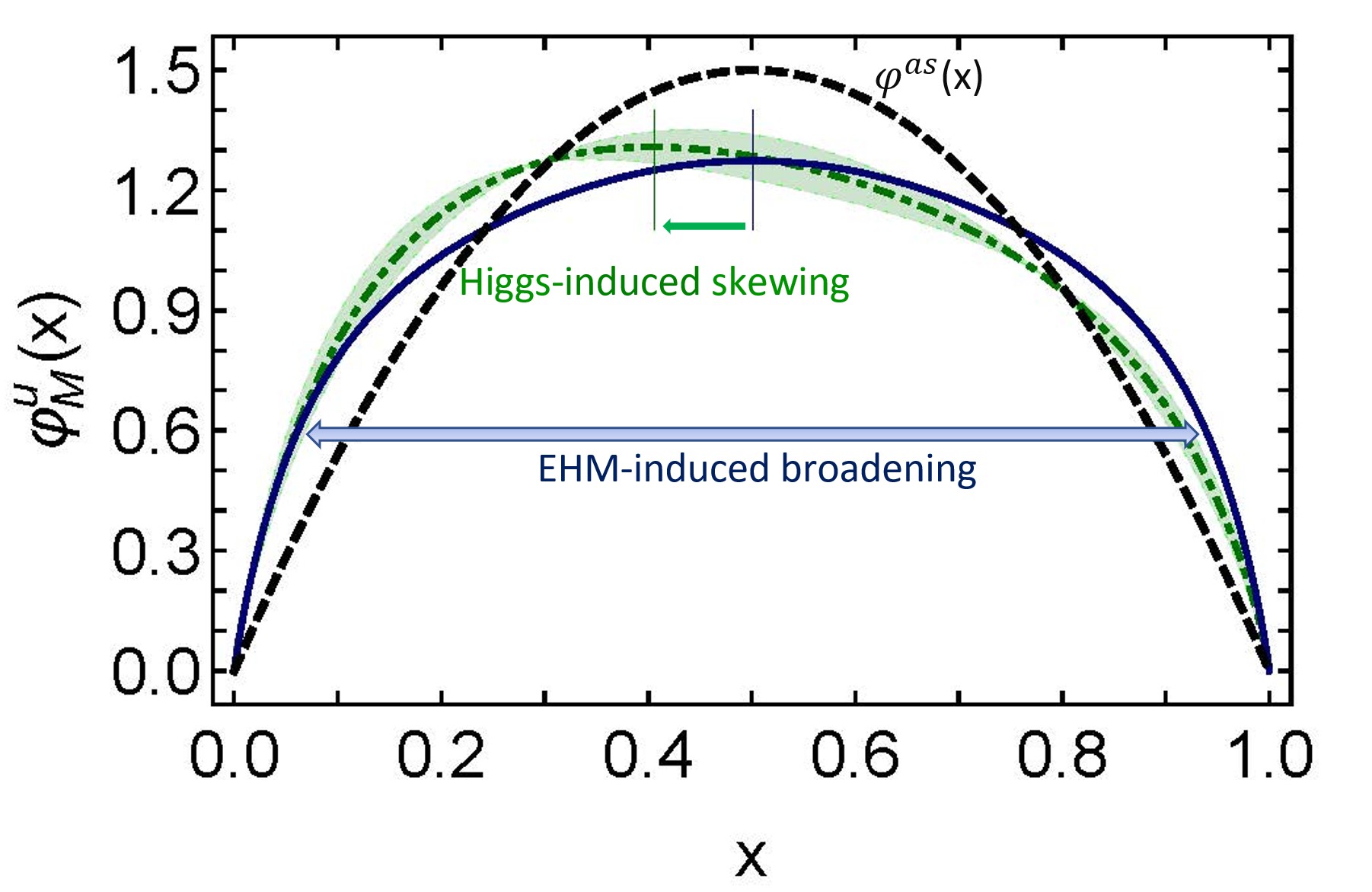}}
\caption{\label{FigPDAs}
DAs charting the light-front momentum distribution of the $u$-quark in a meson $M$: pion, solid blue curve; kaon -- dot-dashed green curve within like-coloured bands; asymptotic DA, $\varphi^{\rm as}(x)=6 x (1-x)$ -- dashed black curve.  All ground-state meson DAs approach $\varphi^{\rm as}(x)$ as $m_p/\zeta \to 0$, where $\zeta$ is the energy scale of the given experiment.  However, at the scales accessible in contemporary experiments, realistic meson DAs are broadened as a consequence of EHM; and in systems defined by valence-quarks with different Higgs-produced current-masses, the peak is shifted away from $x=0.5$.}
\end{figure}

Eq.\,\eqref{EqGTRE} is remarkable and revealing: the former because it is a mathematical statement of equivalence between the one-body and pseudoscalar two-body problems in chiral-limit QCD, problems which are normally considered to be completely independent; and the latter because it states that the cleanest expressions of EHM in the SM are located in the properties of the massless NG modes.  It is worth highlighting here that in the absence of Higgs couplings, all properties of $\pi$- and $K$-mesons are identical.  At realistic Higgs couplings, measurable properties of the $\pi$ and $K$ are windows onto EHM and its modulation by the Higgs boson.  Stated differently, the SM has two mass generating mechanisms and the properties of $\pi$ and $K$ mesons provide clear and direct access to both.

\smallskip

\noindent\emph{6:\;Higgs-modulation of EHM}.\,---\,
Following the advent of quantum mechanics, science has understood that all properties of a bound state are expressed in its wave function.  In relativistic quantum field theory, many appealing features of Schr\"odinger wave functions are preserved if one works with the light-front projections of their covariant analogues \cite{Heinzl:2000ht}.  The simplest such object is a bound-state's leading-twist distribution amplitude (DA), which describes the probability that a given parton carries a fraction $x$ of the meson's total light-front momentum.  At scales $\zeta\gg m_p$, this DA assumes its asymptotic profile \cite{Lepage:1979zb, Efremov:1979qk, Lepage:1980fj}: $\varphi^{\rm as}(x)=6 x (1-x)$.

At scales appropriate to contemporary measurements, the $\pi$ and $K$ DAs have been computed, with the results depicted in Fig.\,\ref{FigPDAs} and discussed in Ref.\,\cite{Cui:2020tdf}.  Two features are readily apparent: the real-world DAs are very different from $\varphi^{\rm as}(x)$ and this is a consequence of EHM \cite{Chang:2013pq}; and the Higgs-generated disparity in size between the current-quark masses of the strange $(s)$ quark and the light $u$, $d$ quarks, which is roughly a factor of 25 \cite{Zyla:2020zbs}, is manifested as merely a 20\% shift in the peak location of the $K$ DA.

In quantum field theory as in quantum mechanics, DAs/wave-functions cannot be directly measured.  However, in terms of the meson's complete leading-twist dressed-parton light-front wave function, $\psi_{M_u}^{\uparrow\downarrow}(x,\vec{k};\zeta_H)$, the meson's DA is obtained as follows:
\begin{equation}
f_M \,\varphi^u_{M}(x;\zeta_H)= \frac{1}{16\pi^3}\int d^2 k_\perp \,\psi_{M_u}^{\uparrow\downarrow}(x,k_\perp^2;\zeta_H)\,,
\end{equation}
where $f_M$ is the meson's leptonic decay constant, with the DA of the partner valence constituent, $\bar h$, obtained via $\varphi_{M}^{\bar h}(x;\zeta_H) =  \varphi_{M}^{u}(1-x;\zeta_H)$.  The related distribution function is defined as \cite{Brodsky:1989pv}
\begin{equation}
\label{PDFLFWF}
{\mathpzc u}^{M}(x;\zeta_H) = \int d^2k_\perp \, |\psi_{M_u}^{\uparrow\downarrow}(x,k_\perp^2;\zeta_H) |^2.
\end{equation}
This quantity, as the modulus-squared of the wave function, is measurable.  Profiting now from the fact that a factorised approximation to $\psi_{M_u}^{\uparrow\downarrow}(x,k_\perp^2;\zeta_H)$ is reliable for integrated quantities when the wave function has fairly uniform support \cite{Xu:2018eii}, one can write
\begin{equation}
\psi_{M_u}^{\uparrow\downarrow}(x,k_\perp^2;\zeta_H) = \varphi_{M}^{u}(x;\zeta_H) \psi_{M_u}^{\uparrow\downarrow}(k_\perp^2;\zeta_H)\,,
\end{equation}
where the optimal choice for $\psi_{M_u}^{\uparrow\downarrow}(k_\perp^2;\zeta_H)$ is determined by the application.  Then
\begin{equation}
\label{PDFeqPDA2}
{\mathpzc u}^{M}(x;\zeta_H) \propto |\varphi_{M}^{u}(x;\zeta_H)|^2,
\end{equation}
with the constant of proportionality fixed by the normalisation condition on $\psi_{M_u}^{\uparrow\downarrow}(x,k_\perp^2;\zeta_H)$.

\begin{figure}[t]
\centerline{\includegraphics[clip, width=0.47\textwidth]{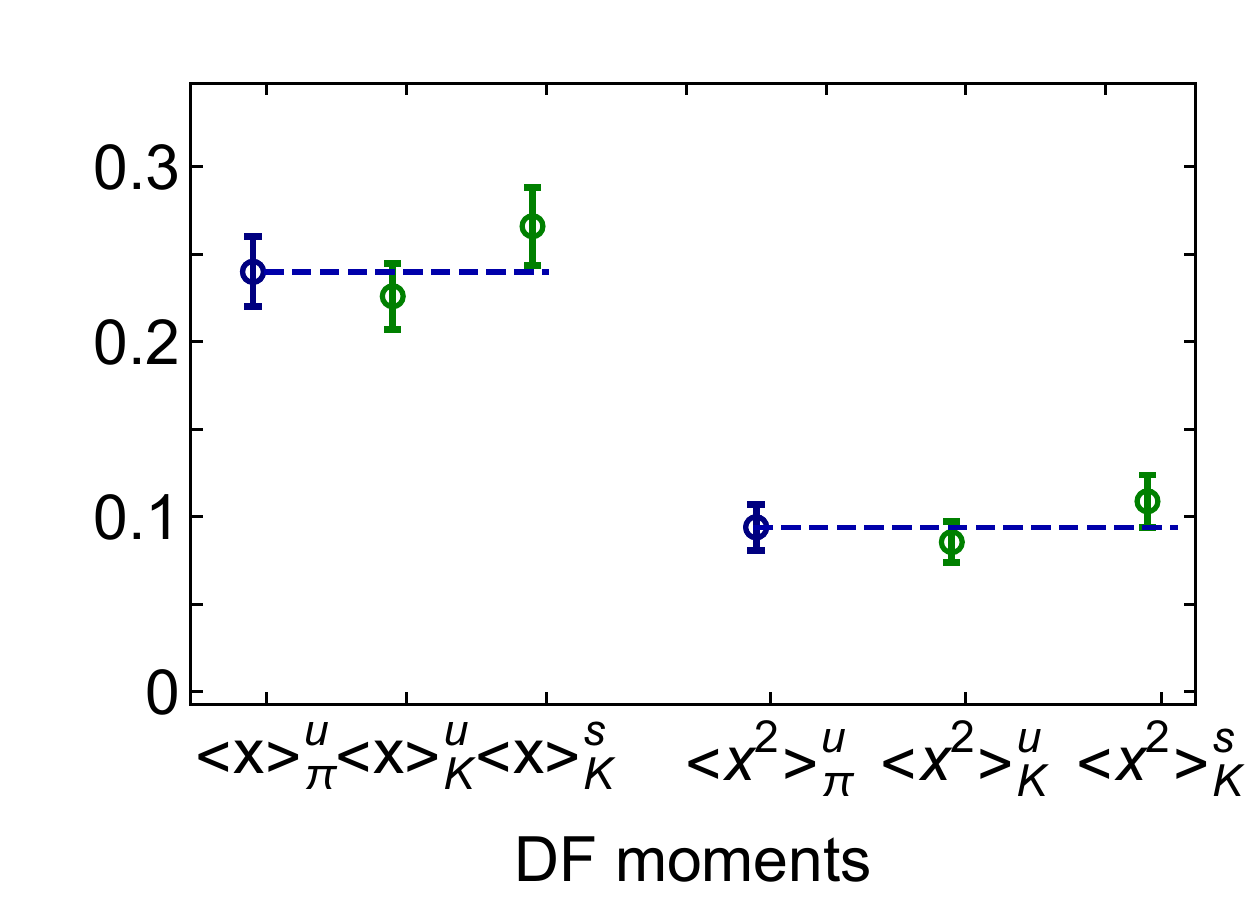}}
\caption{\label{FigMellin}
Low-order $\zeta=\zeta_2$ Mellin moments of the $\pi$ and $K$ DFs, drawn from Table~\ref{moments}.  The horizontal dashed lines are drawn to highlight the pattern of SU$(3)$-flavour symmetry breaking: compared with the $u$-quark-in-$\pi$ values, $\langle x \rangle_K^u$ is reduced by 6\%, $\langle x \rangle_K^{\bar s}$ is increased by 11\% and $\langle x^2 \rangle_K^u$ is reduced by 9\%, $\langle x^2 \rangle_K^{\bar s}$ is increased by 16\%}
\end{figure}

This approach has been used to make the DAs in Fig.\,\ref{FigPDAs} the foundation for predictions of all $\pi$ and $K$ DFs \cite{Cui:2020dlm, Cui:2020tdf}, \emph{i.e}.\ the momentum-fraction probability distributions for valence-quarks, sea-quarks and glue within $\pi$, $K$.  It is thereby enabling measurable connections to be drawn between EHM and Higgs modulation on the one hand and, on the other, the sort of high-energy experiments that first delivered proof for the existence of quarks and gluons \cite{Taylor:1991ew, Kendall:1991np, Friedman:1991nq} and are now being exploited to draw images of hadronic interiors \cite{Accardi:2012qut, Denisov:2018unj, Aguilar:2019teb, Chen:2020ijn}.  Given that the expression of EHM in almost-NG modes differs so greatly from that in the proton (see Fig.\,\ref{F1CDR}), the approach enables one to address many questions of long-standing interest, \emph{e.g}.\ how is the $\pi$-meson's light-front momentum shared amongst its constituents; how are the distributions different in the $K$-meson; and are these in-NG-mode distributions very different from those in the proton?

If one wishes to escape the full complexity of scale evolution in QCD \cite{Dokshitzer:1977sg, Gribov:1972ri, Lipatov:1974qm}, then the valence-quark distributions are best to calculate.  They have been the subject of many studies.  (See, \emph{e.g}.\ Ref.\,\cite{Holt:2010vj} and citations therein and thereto.)  When considering momentum fractions, it is usual to compute Mellin moments of the DFs:
\begin{equation}
\langle x^m \rangle_M^q = \int_{0}^{1}dx\,x^m\,q^M(x;\zeta)\,;
\end{equation}
and typical to quote results at $\zeta=\zeta_2 =2\,$GeV.
Using the DAs in Fig.\,\ref{FigPDAs}, the relation in Eq.\,\eqref{PDFeqPDA2}, and the all-orders $\zeta_H \to \zeta$-evolution procedure explained in Refs.\,\cite{Ding:2019lwe, Cui:2019dwv, Cui:2020dlm, Cui:2020tdf}, one obtains the values of these low order moments depicted in Fig.\,\ref{FigMellin} and listed in Table~\ref{moments}.
\begin{table}[t]
\caption{\label{moments}
Low-order Mellin moments of the $\pi$ and $K$ DFs at $\zeta=\zeta_2=2\,$GeV (Row 1) and $\zeta=\zeta_3=3.1\,$GeV (Row 2), computed following Refs.\,\cite{Cui:2020dlm, Cui:2020tdf}.  The indicated uncertainties express that in the value of $\hat\alpha(0)$ -- see Fig.\,\ref{FigalphaPI}.
(Despite twenty years of improvement in understanding and practice, the $\pi$ results listed here are practically identical to those obtained twenty years ago \cite{Hecht:2000xa}.)}
\begin{tabular*}%{|c|c|c|c|c|c|c|}\hline
{\hsize}
{
l@{\extracolsep{0ptplus1fil}}|
c@{\extracolsep{0ptplus1fil}}
c@{\extracolsep{0ptplus1fil}}
c@{\extracolsep{0ptplus1fil}}
c@{\extracolsep{0ptplus1fil}}
c@{\extracolsep{0ptplus1fil}}
c@{\extracolsep{0ptplus1fil}}}\hline\hline
%
%(\textbf{B})
& $\langle x\rangle_{\pi}^u\ $ & $\langle x \rangle_K^u\ $ & $\langle x \rangle_K^{\bar s}\ $
& $\langle x^2\rangle_{\pi}^u\ $ & $\langle x^2 \rangle_K^u\ $ & $\langle x^2 \rangle_K^{\bar s}\ $ \\\hline
$\zeta_2\ $
& $0.24(2)\ $ & $0.23(2)\ $ & $0.27(2)\ $
& $0.094(13)\ $ & $0.086(12)\ $ & $0.11(2)\phantom{00}\ $\\
$\zeta_3\ $
& $0.22(2)\ $ & $0.21(2)\ $ & $0.25(2)\ $
& $0.085(11)\ $ & $0.078(10)\ $ & $0.097(12)\ $
\\
\hline\hline
\end{tabular*}
\end{table}
%%{{0.9, 0.24, 0.02}, {1.9, 0.226043, 0.0188369}, {2.9, 0.266059, 0.0221716}}
%% {{4.5, 0.094, 0.013}, {5.9, 0.0855239, 0.0118278}, {7.3, 0.108847,  0.0150534}}
%
At the scale $\zeta_2$, valence quarks carry a fraction $0.47(3)$ of the $\pi$-meson's light-front momentum, whereas they carry $0.49(3)$ in the kaon.  That valence-quarks carry more of the kaon's momentum is explained by the fact that $s$-quarks are heavier than $u$- and $d$-quarks.  However, once again, the impact of the Higgs generated current-quark mass differences is very much damped by EHM.

\smallskip

\noindent\emph{7:\;Glue Distributions from $J/\psi$ Production}.\,---\,
Since gluons are electrically inert, direct empirical access to glue distributions in hadrons is difficult if one uses electron/positron beams.  A promising alternative is the study of $J/\psi$-meson production using $\pi$ or $K$ beams incident on a proton target \cite{McEwen:1982fe, Badier:1983dg, Chang:2020rdy} because this process is expected to proceed via gluon+gluon fusion.  The resolving scale relevant to such measurements is $\zeta_3=m_{J/\psi}=3.1\,$GeV; hence in order to assist with future analyses of existing and anticipated data, it is worth calculating the valence, sea and glue distributions in the $\pi$ and $K$ at $\zeta_3$.

\begin{figure}[t]
\vspace*{2ex}

\leftline{\hspace*{0.5em}{\large{\textsf{A}}}}
\vspace*{-5ex}
\includegraphics[clip, width=0.42\textwidth]{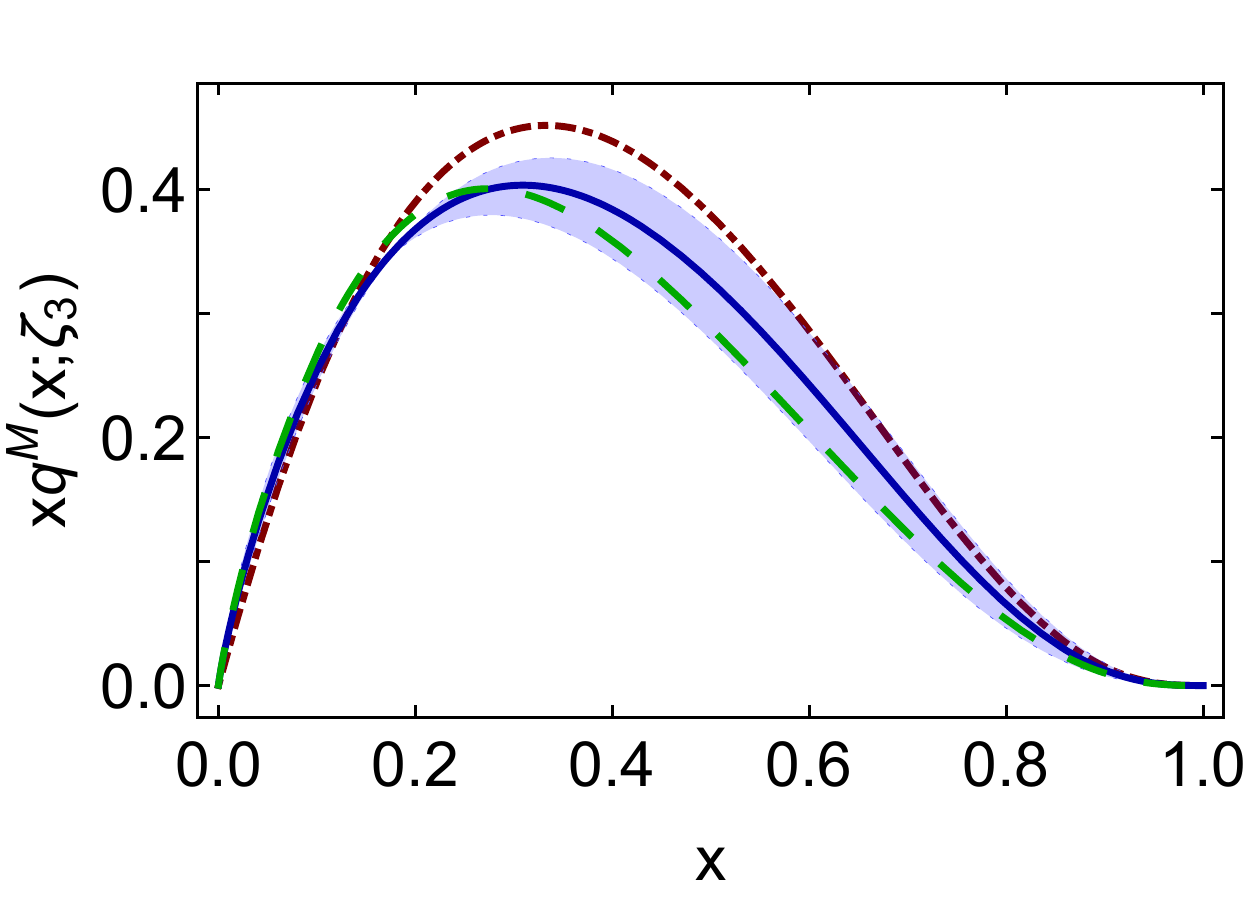}
\vspace*{2ex}

\leftline{\hspace*{0.5em}{\large{\textsf{B}}}}
\vspace*{-5ex}
\includegraphics[clip, width=0.42\textwidth]{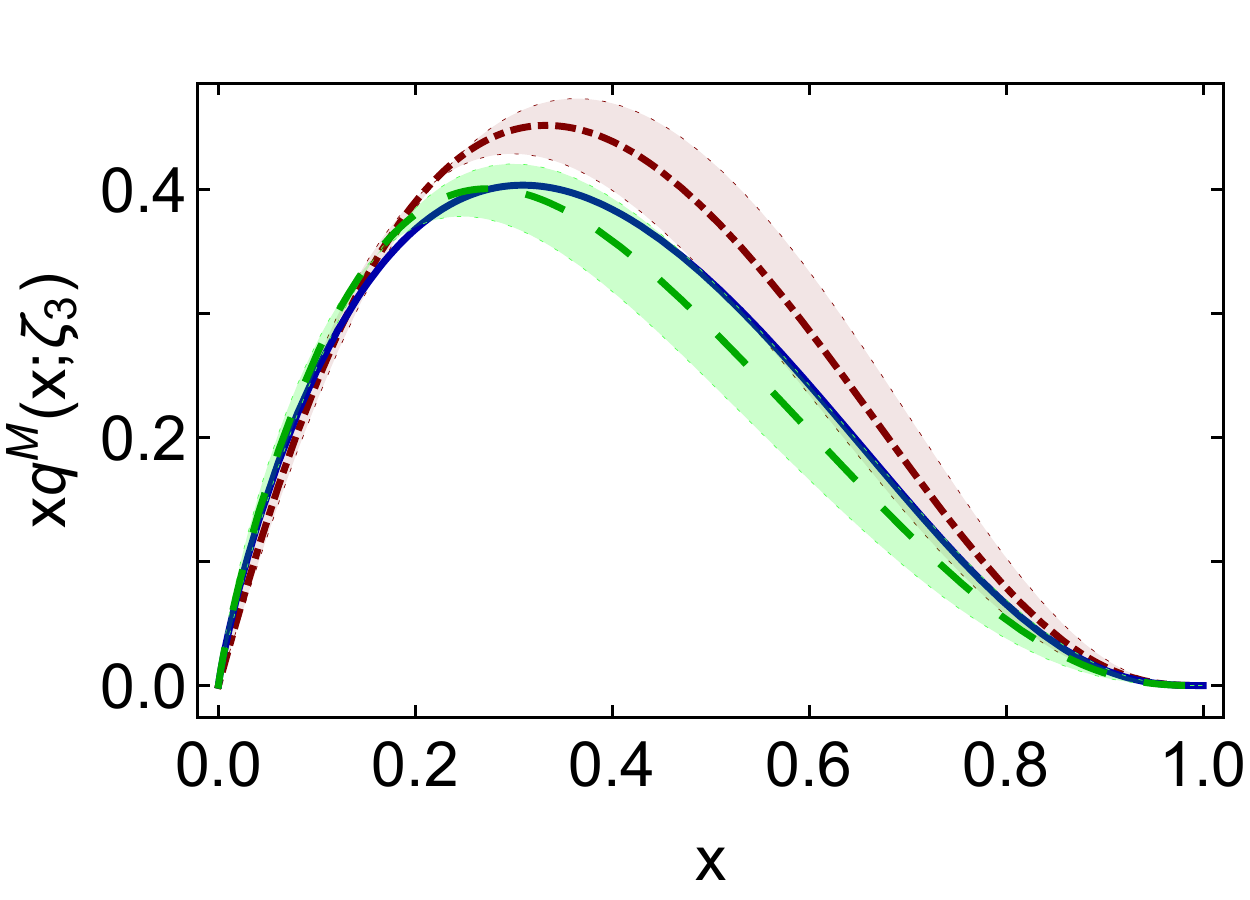}
\caption{\label{FigValence}
Valence-quark distribution functions: $u^\pi$ -- solid blue curve; $u^K$ -- long-dashed green curve; and $\bar s^K$ -- dot-dashed maroon curve.
The upper panel features $u^\pi$, with the blue band indicating the estimated uncertainty owing to that in $\hat\alpha(0)$; whereas the lower panel features $u^K$, $\bar s^K$ with their uncertainties.
(The pointwise form of $x {\mathpzc u}^\pi(x)$ is practically identical to that obtained twenty years ago \cite{Hecht:2000xa} using a more phenomenological approach.)
}
\end{figure}

Using the DAs in Fig.\,\ref{FigPDAs}, Eq.\,\eqref{PDFeqPDA2}, and the all-orders $\zeta$-evolution procedure explained in Refs.\,\cite{Ding:2019lwe, Cui:2019dwv, Cui:2020dlm, Cui:2020tdf}, one obtains the valence quark distributions depicted in Fig.\,\ref{FigValence}.  The curves are interpolated by
\begin{align}
\nonumber & {\mathpzc q}^M(x)  = {\mathpzc n}_{{\mathpzc q}^M} \,x^\alpha (1-x)^\beta \\
& \times [1 + \rho\, x^{\alpha_1/4} (1-x)^{\beta_1/4} + \gamma \,x^{\alpha_1/2} (1-x)^{\beta_1/2} ]\,,
\label{PDFform}
\end{align}
%{\[Alpha] -> -0.10164, \[Beta] ->  2.35649, \[Rho] -> -1.30584, \[Gamma] -> 0.584702}
where ${\mathpzc n}_{{\mathpzc q}^M}$ ensures baryon number conservation and the powers and coefficients are listed in Table~\ref{fittingparametersallV}.  They express the $x\simeq 1$ behaviour predicted by QCD analyses \cite{Farrar:1975yb, Berger:1979du, Brodsky:1994kg} based on the known behavior of hadron wave functions at large valence-quark relative momenta \cite{Lane:1974he, Politzer:1976tv, Lepage:1980fj, Maris:1997hd, Qin:2014vya}.  At $\zeta_3$, the valence degrees-of-freedom carry 45(3)\% of the $\pi$ momentum and 46(3)\% in the $K$.

\begin{table}[t]
\caption{\label{fittingparametersallV}
Coefficients and powers that provide interpolations for the computed valence-quark distribution functions depicted in Fig.\,\ref{FigValence}, when used in Eq.\,\eqref{PDFform}.  The scale is $\zeta_3=3.1\,$GeV.
}
\begin{center}
\begin{tabular*}%{|c|c|c|c|c|c|c|}\hline
{\hsize}
{
l@{\extracolsep{0ptplus1fil}}|
c@{\extracolsep{0ptplus1fil}}
c@{\extracolsep{0ptplus1fil}}
c@{\extracolsep{0ptplus1fil}}
c@{\extracolsep{0ptplus1fil}}
c@{\extracolsep{0ptplus1fil}}
c@{\extracolsep{0ptplus1fil}}
c@{\extracolsep{0ptplus1fil}}}\hline\hline
%
% M {\[Alpha] -> 0.0442687, \[Alpha]1 -> 0.129375, \[Beta]1 ->  0.905553, \[Rho] -> -1.92696, \[Gamma] -> 0.949586}
 ${\mathpzc u}^\pi$
 & ${\mathpzc n}_{{\mathpzc u}^\pi}\ $ & $\alpha\ $ & $\beta\ $ & $\alpha_1\ $ & $\beta_1\ $ & $\rho\ $ & $\gamma\ $ \\\hline
             & $137\phantom{.9}\ $ & $\phantom{-}0.119\phantom{3}\ $ & $3.09\ $ & $0.145\phantom{9}\ $& $0.903\ $& $-1.95\ $ & $0.971\phantom{3}\ $ \\
$\zeta_3\ $ & $118\phantom{.9}\ $ & $\phantom{-}0.0443\ $ & $3.21\ $ & $0.129\phantom{9}\ $& $0.906\ $& $-1.93\ $ & $0.950\phantom{9}\ $ \\
             & $\phantom{1}96.9\ $ & $-0.0450\ $ & $3.35\ $ &$0.109\phantom{9}\ $& $0.911\ $& $-1.90\ $ & $0.925\phantom{9}\ $  \\\hline
%----------
 ${\mathpzc u}^K$
 & ${\mathpzc n}_{{\mathpzc u}^K}\ $ & $\alpha\ $ & $\beta\ $ & $\alpha_1\ $ & $\beta_1\ $ & $\rho\ $ & $\gamma\ $ \\\hline
             & $65.8\ $ & $\phantom{-}0.179\phantom{7}\ $ & $3.09\ $ & $0.358\phantom{9}\ $& $1.39\phantom{3}\ $& $-2.08\ $ & $1.16\phantom{3}\ $ \\
$\zeta_3\ $ & $57.1\ $ & $\phantom{-}0.119\phantom{7}\ $ & $3.21\ $ & $0.375\phantom{9}\ $& $1.46\phantom{3}\ $& $-2.11\ $ & $1.20\phantom{9}\ $ \\
             & $47.4\ $ & $\phantom{-}0.0421\ $ & $3.35\ $ &$0.374\phantom{9}\ $& $1.52\phantom{9}\ $& $-2.11\ $ & $1.21\phantom{9}\ $  \\\hline
$\bar{\mathpzc s}^K$
& ${\mathpzc n}_{\bar {\mathpzc s}^K}\ $ & $\alpha\ $ & $\beta\ $ & $\alpha_1\ $ & $\beta_1\ $ & $\rho\ $ & $\gamma\ $ \\\hline
             & $79.7\ $ & $\phantom{-}0.259\phantom{7}\ $ & $3.03\ $ & $0.235\phantom{5}\ $& $1.39\phantom{9}\ $& $-1.92\phantom{3}\ $ & $0.975\ $ \\
$\zeta_3\ $ & $69.0\ $ & $\phantom{-}0.199\phantom{7}\ $ & $3.14\ $ & $0.228\phantom{5}\ $& $1.39\phantom{9}\ $& $-1.90\phantom{3}\ $ & $0.960\ $ \\
             & $58.8\ $ & $\phantom{-}0.132\phantom{7}\ $ & $3.27\ $ &$0.222\phantom{5}\ $& $1.39\phantom{9}\ $& $-1.89\phantom{3}\ $ & $0.956\ $  \\
\hline\hline
\end{tabular*}
\end{center}
\end{table}

\begin{figure}[t]
\vspace*{2ex}

\leftline{\hspace*{0.5em}{\large{\textsf{A}}}}
\vspace*{-5ex}
\includegraphics[clip, width=0.42\textwidth]{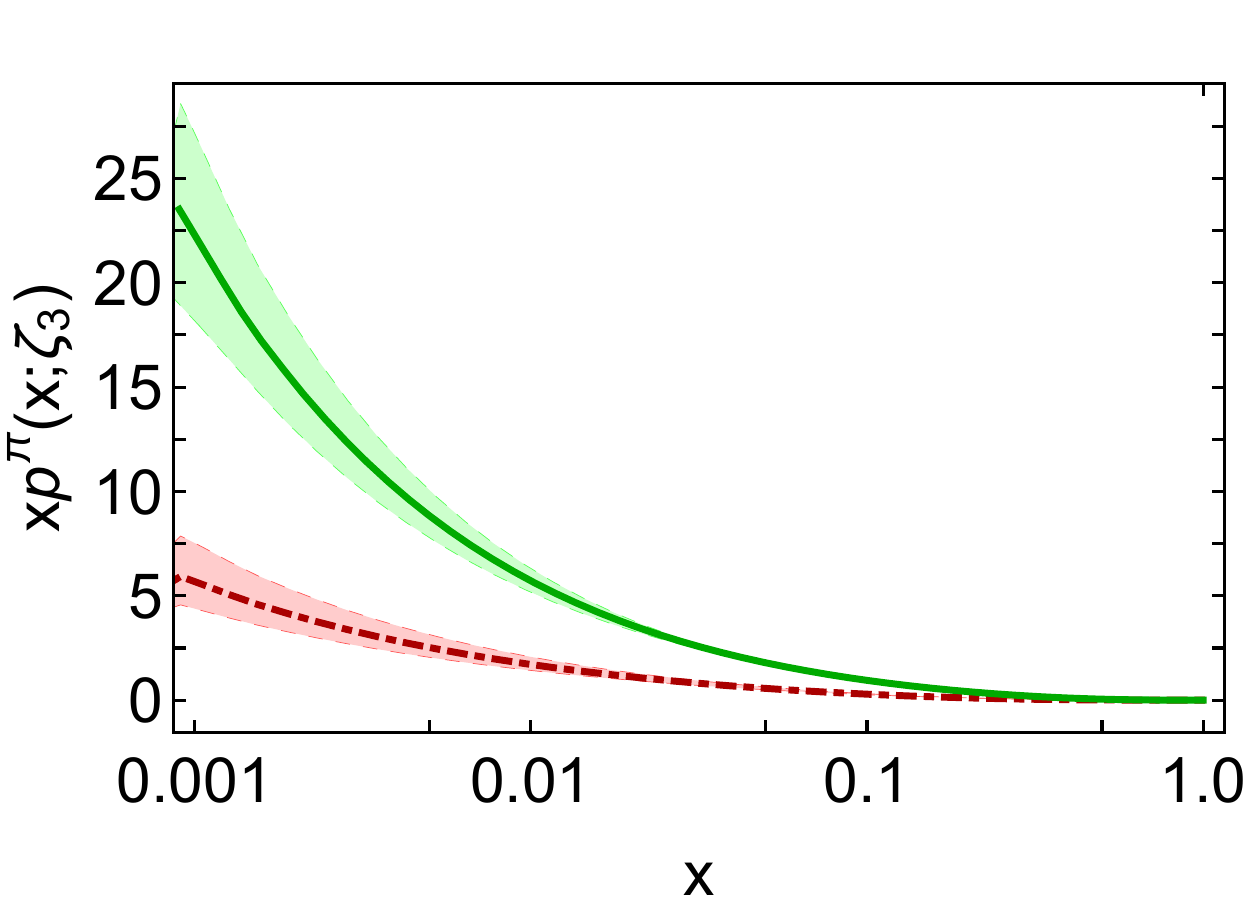}
\vspace*{2ex}

\leftline{\hspace*{0.5em}{\large{\textsf{B}}}}
\vspace*{-5ex}
\includegraphics[clip, width=0.42\textwidth]{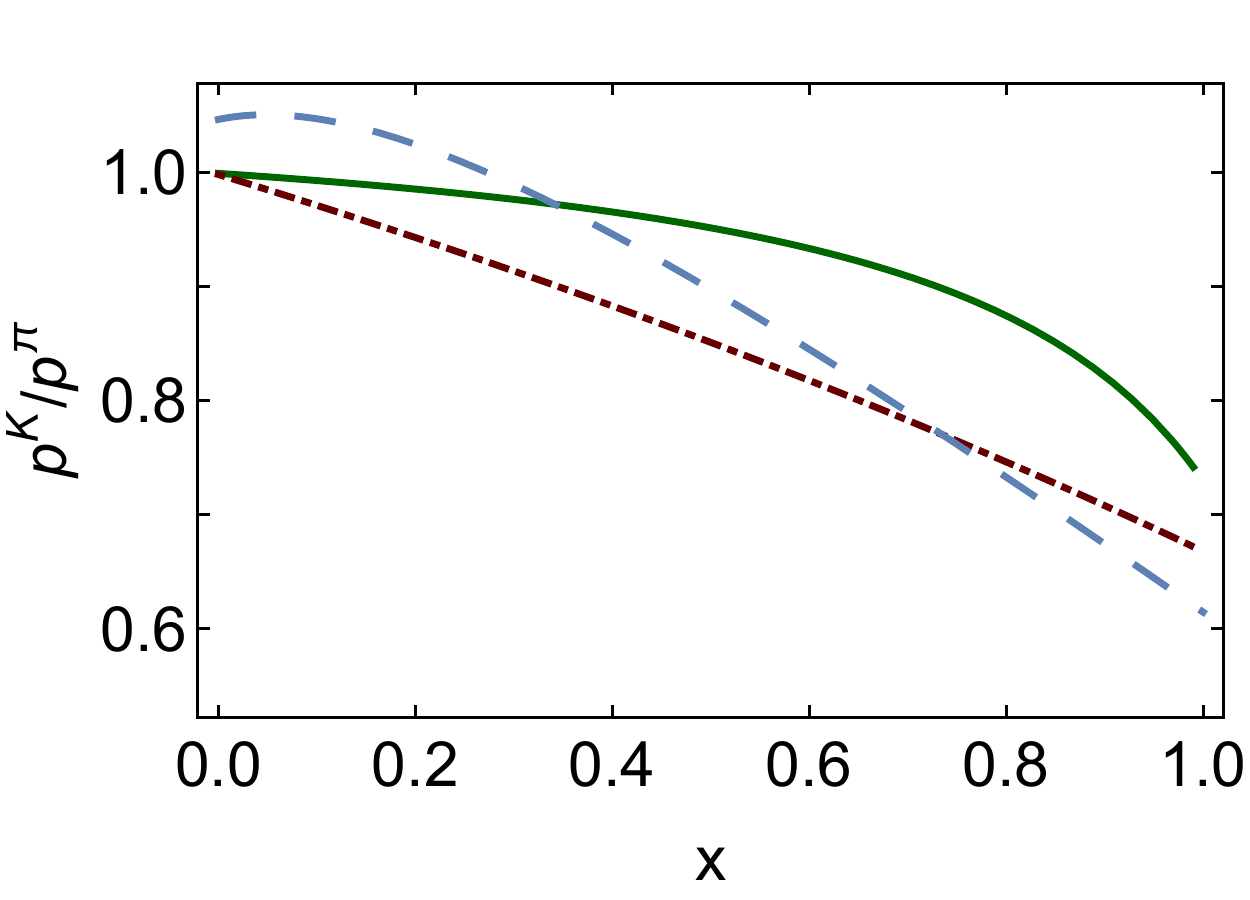}
\caption{\label{FigGSpi}
\emph{Upper panel--A}.  Solid green curve, $p=g$ -- prediction for the pion's glue distribution; and dot-dashed red curve, $p=S$ -- predicted sea-quark distribution.
\emph{Lower panel--B}.  $K/\pi$ DF ratios: $p=g$ -- solid green curve; $p=S$ -- dot-dashed red curve; and $p={\mathpzc u}$ -- long-dashed blue curve.  For comparison, this figure also depicts $u^K(x;\zeta_3)/u^\pi(x;\zeta_3)$.
Normalisation convention: $\langle x[2 {\mathpzc u}^\pi(x;\zeta_3)+g^\pi(x;\zeta_3)+S^\pi(x;\zeta_3)]\rangle=1$.
(The uncertainty bands bracketing the results in Panel\,A reflect the uncertainty  in $\hat\alpha(0)$.  This uncertainty cancels in the ratios depicted in Panel\,B.  Results at $\zeta=3.1\,$GeV.)
}
\end{figure}

Again following Refs.\,\cite{Cui:2020dlm, Cui:2020tdf} and using the valence DFs already presented, the $\zeta=3.1\,$GeV $\pi$-meson glue and sea-quark DFs can be calculated.  The results are depicted in Fig.\,\ref{FigGSpi}A, wherein the curves are effectively interpolated using the following functional form \cite{Gluck:1999xe}:
\begin{equation}
x {\mathpzc p}(x)  = {\mathpzc A} \,x^\alpha (1-x)^\beta  [1 + \rho\, x^{1/2} + \gamma \,x ]\,,
\label{PDFformgS}
\end{equation}
${\mathpzc p} = g, S$, with the coefficients in Table~\ref{GlueSeaCoefficients}.  The associated momentum fractions are ($\zeta=\zeta_3$):
\begin{equation}
\langle x\rangle^\pi_g = 0.43(2)\,, \quad
\langle x\rangle^\pi_{\rm sea} = 0.12(2)\,.
\end{equation}

The $\zeta_3$ glue and sea-quark distributions in the $K$-meson can likewise be obtained.  A good way to describe the predictions is through comparison with the analogous $\pi$-meson results in Fig.\,\ref{FigGSpi}A.  Hence, Fig.\,\ref{FigGSpi}B depicts the following ratios: ${\mathpzc p}^K(x)/{\mathpzc p}^\pi(x)$, ${\mathpzc p}={\mathpzc g},{\mathpzc S}$, which are well described by the following functions:
\begin{equation}
{\mathsf R}_{\mathpzc g}^{K\pi} = \frac{1.00\, -0.842 x}{1-0.786 x} \,, \;
{\mathsf R}_{\mathpzc S}^{K\pi}  = \frac{1.00\, -0.462 x}{1-0.197 x}\,.
\end{equation}
Evidently, the $K$ and $\pi$ glue and sea-quark DFs are quite similar on $x\lesssim 0.2$; but there are noticeable differences on $x\gtrsim 0.2$, \emph{i.e}.\ the domain of valence-quark/antiquark dominance.  These differences are generated by Higgs-boson couplings into QCD and are on the order of $\approx 33$\% at $x=1$ \emph{cf}.: $1-f_\pi^2/f_K^2-1 \approx 0.3$, where $f_M$ is a measure of the size of the meson's wave function at the origin in configuration space; and $1-[M_u(0)/M_s(0)]^2\approx 0.3$ \cite{Cui:2020dlm}, where $M_q(k)$ is the dressed-quark mass function, whose chiral limit form is drawn in Fig.\,\ref{Figmk}.

\begin{table}[t]
\caption{\label{GlueSeaCoefficients}
Coefficients and powers that reproduce the computed pion's glue and sea-quark distribution functions depicted in Fig.~\ref{FigGSpi}A when used in Eq.\,\eqref{PDFformgS}.
}
\begin{center}
\begin{tabular*}%{|c|c|c|c|c|c|c|}\hline
{\hsize}
{
l@{\extracolsep{0ptplus1fil}}|
c@{\extracolsep{0ptplus1fil}}
c@{\extracolsep{0ptplus1fil}}
c@{\extracolsep{0ptplus1fil}}
c@{\extracolsep{0ptplus1fil}}
c@{\extracolsep{0ptplus1fil}}}\hline\hline
 $\zeta_3\ $ & ${\mathpzc A}\ $ & $\alpha\ $ & $\beta\ $ & $\rho\ $ & $\gamma\ $ \\\hline
             % & $10.98\ $ & $-0.052\ $ & $2.29\ $ & $-1.40\ $ & $0.637\ $  \\
             & $0.462\ $ & $-0.539\ $ & $4.09\ $ & $-0.296\ $& $\phantom{-}0.229\ $\\
%$\zeta_2\ $ & $\phantom{1}9.26\ $ & $-0.096\ $ & $2.37\ $ & $-1.32\ $ & $0.594\ $ \\
  $g\ $ & $0.735\ $ & $-0.494\ $ & $4.21\ $ & $-1.54\phantom{1}\ $ & $\phantom{-}1.36\phantom{9}\ $\\
            % & $\phantom{1}7.38\ $ & $-0.140\ $ & $2.44\ $ & $-1.21\ $ & $0.538\ $ \\\hline
             & $0.295\ $ & $-0.638\ $ & $4.35\ $ &$\phantom{-}2.23\phantom{1}\ $& $-5.08\phantom{9}\ $\\\hline
             & $0.144\ $ & $-0.488\ $ & $5.09\ $ & $\phantom{-}0.956\ $& $-2.36\phantom{9}\ $\\
%$\zeta_2\ $ & $\phantom{1}9.26\ $ & $-0.096\ $ & $2.37\ $ & $-1.32\ $ & $0.594\ $ \\
$S\ $ & $0.127\ $ & $-0.538\ $ & $5.21\ $ & $\phantom{-}2.20\phantom{8}\ $ & $-4.82\phantom{9}\ $\\
            % & $\phantom{1}7.38\ $ & $-0.140\ $ & $2.44\ $ & $-1.21\ $ & $0.538\ $ \\\hline
             & $0.108\ $ & $-0.595\ $ & $5.35\ $ &$\phantom{-}3.54\phantom{8}\ $& $-7.50\phantom{9}\ $\\
\hline\hline
\end{tabular*}
\end{center}
\end{table}

\smallskip

\noindent\emph{8:\;Epilogue}.\,---\,
Ren\'e Descartes, the 17th century mathematician and philosopher, is widely celebrated for introducing the notion that all which is not human is merely the sum of its parts.  This first statement of a \emph{reductionist} position is largely held to be the foundation for modern science.  It leads to the question posed at the outset; namely: Is there a Lagrangian for Nature, ${\mathpzc L}_{N}$?  If so, then it must contain and define the natural mass (length) scale for all materials.  Further, if ${\mathpzc L}_{N}$ exists, then logic, as expressed in mathematics, is not something invented to describe Nature; instead, it is innate to Nature or even the essence of Nature.  If this is true, then there are answers to all questions that have been asked and also to those questions that have not yet arisen.  This is worth considering.

The far lesser issues discussed herein also focus on mass: why is the proton mass roughly 2\,000-times the electron mass; and why is its cousin, the $\pi$-meson, so much lighter in comparison?  An intimately related question was not explicitly addressed, \emph{viz}.\ why is the proton absolutely stable?  This relates to the question of gluon and quark confinement; and the perspective related herein links confinement directly with the emergence of hadronic mass.  As detailed elsewhere \cite{Roberts:2020hiw}, the dynamical generation of nuclear-size gluon and quark masses and their associated running mass functions is necessary and sufficient for a dynamical realisation of confinement.

The current paradigm for addressing this array of questions is quantum chromodynamics (QCD).  As sketched above, QCD is plausibly (probably?) a mathematically well-defined quantum field theory in four spacetime dimensions, the only such theory science has ever produced.  Consequently, it can potentially serve as an archetype for extending the Standard Model to cover those perceived phenomena which physics cannot yet explain.

Although QCD is defined by a seemingly simple Lagrangian, it specifies a problem that has defied solution for more than forty years.  The key challenges in modern nuclear and high-energy physics are to reveal the observable content of strong QCD and, ultimately, therefrom derive the properties of nuclei.  That progress which has already been made was delivered by an amalgam of experiment, phenomenology, and theory.  The successes have inspired the construction and planning of new-generation facilities.  Science will only profit from these investments if existing synergies between those three branches are exploited and expanded.

\smallskip

\noindent\emph{Acknowledgements}.
These remarks are based on results obtained and insights developed through collaborations with
many people, to all of whom I am greatly indebted.

%%\bibliographystyle{../../../../../zProc/z10/z10KITPC/h-physrev4}
%%\bibliography{../../../../../CollectedBiB}

\begin{thebibliography}{100}

\bibitem{Ellis:2014sjr}
G.~Ellis and J.~Silk,
\newblock Nature {\bf 516}, 321 (2014).

\bibitem{Aad:2012tfa}
G.~Aad {\em et~al.},
\newblock Phys. Lett. B {\bf 716}, 1 (2012).
%%CITATION = ARXIV:1207.7214;%%

\bibitem{Chatrchyan:2012xdj}
S.~Chatrchyan {\em et~al.},
\newblock Phys. Lett. B {\bf 716}, 30 (2012).
%%CITATION = ARXIV:1207.7235;%%

\bibitem{Englert:2014zpa}
F.~Englert,
\newblock Rev. Mod. Phys. {\bf 86}, 843 (2014).
%%CITATION = RMPHA,86,843;%%

\bibitem{Higgs:2014aqa}
P.~W. Higgs,
\newblock Rev. Mod. Phys. {\bf 86}, 851 (2014).
%%CITATION = RMPHA,86,851;%%

\bibitem{Schwinger:1962tp}
J.~S. Schwinger,
\newblock Phys. Rev. {\bf 128}, 2425 (1962).

\bibitem{Zyla:2020zbs}
P.~Zyla {\em et~al.},
\newblock PTEP {\bf 2020}, 083C01 (2020).

\bibitem{Marciano:1979wa}
W.~J. Marciano and H.~Pagels,
\newblock Nature {\bf 279}, 479 (1979).
%%CITATION = NATUA,279,479;%%

\bibitem{Marciano:1977su}
W.~J. Marciano and H.~Pagels,
\newblock Phys. Rept. {\bf 36}, 137 (1978).
%%CITATION = PRPLC,36,137;%%

\bibitem{IZ80}
C.~Itzykson and J.-B. Zuber,
\newblock {\em Quantum Field Theory} (McGraw-Hill Inc., New York, 1980).

\bibitem{Rakow:1990jv}
P.~E.~L. Rakow,
\newblock Nucl. Phys. B {\bf 356}, 27 (1991).
%%CITATION = NUPHA,B356,27;%%

\bibitem{Reenders:1999bg}
M.~Reenders,
\newblock Phys. Rev. D {\bf 62}, 025001 (2000).
%%CITATION = HEP-TH/9908158;%%

\bibitem{Kizilersu:2014ela}
A.~K{\i}z{\i}lers{\"u}, T.~Sizer, M.~R. Pennington, A.~G. Williams and
  R.~Williams,
\newblock Phys. Rev. D {\bf 91}, 065015 (2015).
%%CITATION = ARXIV:1409.5979;%%

\bibitem{Politzer:2005kc}
H.~D. Politzer,
\newblock Proc. Nat. Acad. Sci. {\bf 102}, 7789 (2005).
%%CITATION = PNASA,102,7789;%%

\bibitem{Wilczek:2005az}
F.~Wilczek,
\newblock Proc. Nat. Acad. Sci. {\bf 102}, 8403 (2005).
%%CITATION = HEP-PH/0502113;%%

\bibitem{Gross:2005kv}
D.~J. Gross,
\newblock Proc. Nat. Acad. Sci. {\bf 102}, 9099 (2005).
%%CITATION = PNASA,102,9099;%%

\bibitem{millennium:2006}
{\emph{The Millenium Prize Problems}, eds. J. Carlson, A. Jaffe, and A. Wiles.
  (American Mathematical Society, Providence, 2006)}.

\bibitem{Cornwall:1981zr}
J.~M. Cornwall,
\newblock Phys. Rev. D {\bf 26}, 1453 (1982).
%%CITATION = PHRVA,D26,1453;%%

\bibitem{Taylor:1971ff}
J.~C. Taylor,
\newblock Nucl. Phys. B {\bf 33}, 436 (1971).
%%CITATION = NUPHA,B33,436;%%

\bibitem{Slavnov:1972fg}
A.~A. Slavnov,
\newblock Theor. Math. Phys. {\bf 10}, 99 (1972).
%%CITATION = TMPHA,10,99;%%

\bibitem{Roberts:2016vyn}
C.~D. Roberts,
\newblock Few Body Syst. {\bf 58}, 5 (2017).
%%CITATION = ARXIV:1606.03909;%%

\bibitem{tarrach}
P.~Pascual and R.~Tarrach,
\newblock {\em QCD: Renormalization for the Practitioner} (Springer-Verlag,
  Berlin, 1984),
\newblock {L}ecture Notes in Physics \textbf{194}.

\bibitem{Cui:2019dwv}
Z.-F. Cui {\em et~al.},
\newblock Chin. Phys. C {\bf 44}, 083102 (2020).
%%CITATION = ARXIV:1912.08232;%%

\bibitem{Blum:2014tka}
T.~Blum {\em et~al.},
\newblock Phys. Rev. D {\bf 93}, 074505 (2016).
%%CITATION = ARXIV:1411.7017;%%

\bibitem{Boyle:2015exm}
P.~A. Boyle {\em et~al.},
\newblock Phys. Rev. D {\bf 93}, 054502 (2016).
%%CITATION = ARXIV:1511.01950;%%

\bibitem{Boyle:2017jwu}
P.~A. Boyle {\em et~al.},
\newblock JHEP {\bf 12}, 008 (2017).
%%CITATION = ARXIV:1701.02644;%%

\bibitem{Zafeiropoulos:2019flq}
S.~Zafeiropoulos, P.~Boucaud, F.~De~Soto, J.~Rodr{\'{\i}}guez-Quintero and
  J.~Segovia,
\newblock Phys. Rev. Lett. {\bf 122}, 162002 (2019).
%%CITATION = ARXIV:1902.08148;%%

\bibitem{Aguilar:2019uob}
A.~Aguilar {\em et~al.},
\newblock Eur. Phys. J. C {\bf 80}, 154 (2020).

\bibitem{Chang:2011ei}
L.~Chang and C.~D. Roberts,
\newblock Phys. Rev. C {\bf 85}, 052201(R) (2012).
%%CITATION = 1104.4821;%%

\bibitem{Chang:2013pq}
L.~Chang {\em et~al.},
\newblock Phys. Rev. Lett. {\bf 110}, 132001 (2013).
%%CITATION = ARXIV:1301.0324;%%

\bibitem{Chang:2013epa}
L.~Chang, C.~D. Roberts and S.~M. Schmidt,
\newblock Phys. Lett. B {\bf 727}, 255 (2013).
%%CITATION = ARXIV:1308.4708;%%

\bibitem{Binosi:2016wcx}
D.~Binosi, L.~Chang, J.~Papavassiliou, S.-X. Qin and C.~D. Roberts,
\newblock Phys. Rev. D {\bf 95}, 031501(R) (2017).
%%CITATION = ARXIV:1609.02568;%%

\bibitem{Binosi:2014aea}
D.~Binosi, L.~Chang, J.~Papavassiliou and C.~D. Roberts,
\newblock Phys. Lett. B {\bf 742}, 183 (2015).
%%CITATION = ARXIV:1412.4782;%%

\bibitem{Aguilar:2015bud}
A.~C. Aguilar, D.~Binosi and J.~Papavassiliou,
\newblock Front. Phys. China {\bf 11}, 111203 (2016).
%%CITATION = ARXIV:1511.08361;%%

\bibitem{Gao:2017uox}
F.~Gao, S.-X. Qin, C.~D. Roberts and J.~Rodr{\'{\i}}guez-Quintero,
\newblock Phys. Rev. D {\bf 97}, 034010 (2018).
%%CITATION = ARXIV:1706.04681;%%

\bibitem{Huber:2018ned}
M.~Q. Huber,
\newblock Phys. Rept. {\bf 879}, 1  (2020).

\bibitem{Fischer:2018sdj}
C.~S. Fischer,
\newblock Prog. Part. Nucl. Phys. {\bf 105}, 1 (2019).
%%CITATION = ARXIV:1810.12938;%%

\bibitem{Roberts:2020hiw}
C.~D. Roberts,
\newblock Symmetry {\bf 12}, 1468 (2020).

\bibitem{Qin:2020rad}
S.-X. Qin and C.~D. Roberts,
\newblock Chin. Phys. Lett. {\bf 37}, 121201 (2020).

\bibitem{Binosi:2016nme}
D.~Binosi, C.~Mezrag, J.~Papavassiliou, C.~D. Roberts and
  J.~Rodr{\'i}guez-Quintero,
\newblock Phys. Rev. D {\bf 96}, 054026 (2017).
%%CITATION = ARXIV:1612.04835;%%

\bibitem{GellMann:1954fq}
M.~Gell-Mann and F.~E. Low,
\newblock Phys. Rev. {\bf 95}, 1300 (1954).
%%CITATION = PHRVA,95,1300;%%

\bibitem{Deur:2016tte}
A.~Deur, S.~J. Brodsky and G.~F. de~Teramond,
\newblock Prog. Part. Nucl. Phys. {\bf 90}, 1 (2016).
%%CITATION = ARXIV:1604.08082;%%

\bibitem{RuizdeElvira:2017stg}
J.~Ruiz~de Elvira, M.~Hoferichter, B.~Kubis and U.-G. Mei\ss{}ner,
\newblock J. Phys. G {\bf 45}, 024001 (2018).

\bibitem{Ding:2019qlr}
M.~Ding {\em et~al.},
\newblock Chin. Phys. C (Lett.) {\bf 44}, 031002 (2020).
%%CITATION = ARXIV:1912.07529;%%

\bibitem{Ding:2019lwe}
M.~Ding {\em et~al.},
\newblock Phys. Rev. D {\bf 101}, 054014 (2020).

\bibitem{Cui:2020dlm}
Z.-F. Cui {\em et~al.},
\newblock Eur. Phys. J. A (Lett.) {\bf 57}, 5 (2021).

\bibitem{Cui:2020tdf}
Z.-F. Cui {\em et~al.},
\newblock Eur. Phys. J. C {\bf 80}, 1064 (2020).

\bibitem{Grunberg:1982fw}
G.~Grunberg,
\newblock Phys. Rev. D {\bf 29}, 2315 (1984).
%%CITATION = PHRVA,D29,2315;%%

\bibitem{Dokshitzer:1998nz}
Y.~L. Dokshitzer,
\newblock {\emph{Perturbative QCD theory (includes our knowledge of
  \mbox{$\alpha(s)$})} - hep-ph/9812252},
\newblock in {\em {High-energy physics. Proceedings, 29th International
  Conference, ICHEP'98, Vancouver, Canada, July 23-29, 1998. Vol. 1, 2}}, pp.
  305--324, 1998.
%%CITATION = HEP-PH/9812252;%%

\bibitem{GellMann:1964nj}
M.~Gell-Mann,
\newblock Phys. Lett. {\bf 8}, 214 (1964).
%%CITATION = PHLTA,8,214;%%

\bibitem{Zweig:1981pd}
G.~Zweig,
\newblock (1964),
\newblock {\emph{An $SU(3)$ model for strong interaction symmetry and its
  breaking. Parts 1 and 2} (CERN Reports No.\ 8182/TH.\ 401 and No.\ 8419/TH.\
  412)}.
%%CITATION = CERN-TH-401 ETC.;%%

\bibitem{Lane:1974he}
K.~D. Lane,
\newblock Phys. Rev. D {\bf 10}, 2605 (1974).
%%CITATION = PHRVA,D10,2605;%%

\bibitem{Politzer:1976tv}
H.~D. Politzer,
\newblock Nucl. Phys. B {\bf 117}, 397 (1976).
%%CITATION = NUPHA,B117,397;%%

\bibitem{Dyson:1949ha}
F.~J. Dyson,
\newblock Phys. Rev. {\bf 75}, 1736 (1949).
%%CITATION = PHRVA,75,1736;%%

\bibitem{Bhagwat:2003vw}
M.~S. Bhagwat, M.~A. Pichowsky, C.~D. Roberts and P.~C. Tandy,
\newblock Phys. Rev. C {\bf 68}, 015203 (2003).
%%CITATION = NUCL-TH/0304003;%%

\bibitem{Bowman:2005vx}
P.~O. Bowman {\em et~al.},
\newblock Phys. Rev. D {\bf 71}, 054507 (2005).
%%CITATION = HEP-LAT/0501019;%%

\bibitem{Bhagwat:2006tu}
M.~S. Bhagwat and P.~C. Tandy,
\newblock AIP Conf. Proc. {\bf 842}, 225 (2006).
%%CITATION = NUCL-TH/0601020;%%

\bibitem{Bermudez:2017bpx}
R.~Bermudez, L.~Albino, L.~X. Guti{\'e}rrez-Guerrero, M.~E. Tejeda-Yeomans and
  A.~Bashir,
\newblock Phys. Rev. D {\bf 95}, 034041 (2017).
%%CITATION = ARXIV:1702.04437;%%

\bibitem{Aguilar:2018epe}
A.~C. Aguilar, J.~C. Cardona, M.~N. Ferreira and J.~Papavassiliou,
\newblock Phys. Rev. D {\bf 98}, 014002 (2018).
%%CITATION = ARXIV:1804.04229;%%

\bibitem{Oliveira:2020yac}
O.~Oliveira, T.~Frederico and W.~de~Paula,
\newblock Eur. Phys. J. C {\bf 80}, 484 (2020).

\bibitem{Ivanov:1998ms}
M.~A. Ivanov, {\mbox{Yu}}.~L. Kalinovsky and C.~D. Roberts,
\newblock Phys. Rev. D {\bf 60}, 034018 (1999).
%%CITATION = NUCL-TH/9812063;%%

\bibitem{Qin:2019hgk}
S.-X. Qin, C.~D. Roberts and S.~M. Schmidt,
\newblock Few Body Syst. {\bf 60}, 26 (2019).
%%CITATION = ARXIV:1902.00026;%%

\bibitem{Yin:2019bxe}
P.-L. Yin {\em et~al.},
\newblock Phys. Rev. D {\bf 100}, 034008 (2019).
%%CITATION = ARXIV:1903.00160;%%

\bibitem{Gutierrez-Guerrero:2019uwa}
L.~Guti{\'e}rrez-Guerrero, A.~Bashir, M.~A. Bedolla and E.~Santopinto,
\newblock Phys. Rev. D {\bf 100}, 114032 (2019).

\bibitem{Brodsky:2020vco}
S.~J. Brodsky {\em et~al.},
\newblock Intern. J. Mod. Phys. E {\bf 124}, 2030006 (2020).

\bibitem{Carman:2020qmb}
D.~Carman, K.~Joo and V.~Mokeev,
\newblock Few Body Syst. {\bf 61}, 29 (2020).

\bibitem{Barabanov:2020jvn}
{Barabanov, M. Yu; others},
\newblock Prog. Part. Nucl. Phys. {\bf 116}, 103835 (2021).

\bibitem{Ding:2018xwy}
M.~Ding {\em et~al.},
\newblock Phys. Rev. D {\bf 99}, 014014 (2019).
%%CITATION = ARXIV:1810.12313;%%

\bibitem{Lan:2019rba}
J.~Lan, C.~Mondal, S.~Jia, X.~Zhao and J.~P. Vary,
\newblock Phys. Rev. D {\bf 101}, 034024 (2020).

\bibitem{Chang:2020kjj}
L.~Chang, K.~Raya and X.~Wang,
\newblock Chin. Phys. C {\bf 44}, 114105 (2020).

\bibitem{Kock:2020frx}
A.~Kock, Y.~Liu and I.~Zahed,
\newblock Phys. Rev. D {\bf 102}, 014039 (2020).

\bibitem{Christos:1984tu}
G.~A. Christos,
\newblock Phys. Rept. {\bf 116}, 251 (1984).
%%CITATION = PRPLC,116,251;%%

\bibitem{Roberts:2020udq}
C.~D. Roberts and S.~M. Schmidt,
\newblock Eur. Phys. J. ST {\bf 229}, 3319 (2020).

\bibitem{Maris:1997hd}
P.~Maris, C.~D. Roberts and P.~C. Tandy,
\newblock Phys. Lett. B {\bf 420}, 267 (1998).
%%CITATION = NUCL-TH/9707003;%%

\bibitem{Qin:2014vya}
S.-X. Qin, C.~D. Roberts and S.~M. Schmidt,
\newblock Phys. Lett. B {\bf 733}, 202 (2014).
%%CITATION = ARXIV:1402.1176;%%

\bibitem{Heinzl:2000ht}
T.~Heinzl,
\newblock Lect. Notes Phys. {\bf 572}, 55 (2001).

\bibitem{Lepage:1979zb}
G.~P. Lepage and S.~J. Brodsky,
\newblock Phys. Lett. B {\bf 87}, 359 (1979).
%%CITATION = PHLTA,B87,359;%%

\bibitem{Efremov:1979qk}
A.~V. Efremov and A.~V. Radyushkin,
\newblock Phys. Lett. B {\bf 94}, 245 (1980).
%%CITATION = PHLTA,B94,245;%%

\bibitem{Lepage:1980fj}
G.~P. Lepage and S.~J. Brodsky,
\newblock Phys. Rev. D {\bf 22}, 2157 (1980).
%%CITATION = PHRVA,D22,2157;%%

\bibitem{Brodsky:1989pv}
S.~J. Brodsky and G.~P. Lepage,
\newblock Adv. Ser. Direct. High Energy Phys. {\bf 5}, 93 (1989).
%%CITATION = 00319,5,93;%%

\bibitem{Xu:2018eii}
S.-S. Xu, L.~Chang, C.~D. Roberts and H.-S. Zong,
\newblock Phys. Rev. D {\bf 97}, 094014 (2018).
%%CITATION = ARXIV:1802.09552;%%

\bibitem{Taylor:1991ew}
R.~E. Taylor,
\newblock Rev. Mod. Phys. {\bf 63}, 573 (1991).
%%CITATION = RMPHA,63,573;%%

\bibitem{Kendall:1991np}
H.~W. Kendall,
\newblock Rev. Mod. Phys. {\bf 63}, 597 (1991).
%%CITATION = RMPHA,63,597;%%

\bibitem{Friedman:1991nq}
J.~I. Friedman,
\newblock Rev. Mod. Phys. {\bf 63}, 615 (1991).
%%CITATION = RMPHA,63,615;%%

\bibitem{Accardi:2012qut}
A.~Accardi {\em et~al.},
\newblock Eur. Phys. J. A {\bf 52}, 268 (2016).
%%CITATION = ARXIV:1212.1701;%%

\bibitem{Denisov:2018unj}
O.~Denisov {\em et~al.},
\newblock {\emph{Letter of Intent (Draft 2.0): A New QCD facility at the M2
  beam line of the CERN SPS} -- arXiv:1808.00848 [hep-ex]}.
%%CITATION = ARXIV:1808.00848;%%

\bibitem{Aguilar:2019teb}
A.~C. Aguilar {\em et~al.},
\newblock Eur. Phys. J. A {\bf 55}, 190 (2019).
%%CITATION = ARXIV:1907.08218;%%

\bibitem{Chen:2020ijn}
X.~Chen, F.-K. Guo, C.~D. Roberts and R.~Wang,
\newblock Few Body Syst. {\bf 61}, 43 (2020).

\bibitem{Dokshitzer:1977sg}
Y.~L. Dokshitzer,
\newblock Sov. Phys. JETP {\bf 46}, 641 (1977).
%%CITATION = SPHJA,46,641;%%

\bibitem{Gribov:1972ri}
V.~Gribov and L.~Lipatov,
\newblock Sov. J. Nucl. Phys. {\bf 15}, 438 (1972).

\bibitem{Lipatov:1974qm}
L.~N. Lipatov,
\newblock Sov. J. Nucl. Phys. {\bf 20}, 94 (1975).
%%CITATION = SJNCA,20,94;%%

\bibitem{Holt:2010vj}
R.~J. Holt and C.~D. Roberts,
\newblock Rev. Mod. Phys. {\bf 82}, 2991 (2010).
%%CITATION = 1002.4666;%%

\bibitem{Hecht:2000xa}
M.~B. Hecht, C.~D. Roberts and S.~M. Schmidt,
\newblock Phys. Rev. C {\bf 63}, 025213 (2001).
%%CITATION = NUCL-TH/0008049;%%

\bibitem{McEwen:1982fe}
J.~G. McEwen {\em et~al.},
\newblock Phys. Lett. B {\bf 121}, 198 (1983).
%%CITATION = PHLTA,121B,198;%%

\bibitem{Badier:1983dg}
J.~Badier {\em et~al.},
\newblock Z. Phys. C {\bf 20}, 101 (1983).
%%CITATION = ZEPYA,C20,101;%%

\bibitem{Chang:2020rdy}
W.-C. Chang, J.-C. Peng, S.~Platchkov and T.~Sawada,
\newblock Phys. Rev. D {\bf 102}, 054024 (2020).

\bibitem{Farrar:1975yb}
G.~R. Farrar and D.~R. Jackson,
\newblock Phys. Rev. Lett. {\bf 35}, 1416 (1975).
%%CITATION = PRLTA,35,1416;%%

\bibitem{Berger:1979du}
E.~L. Berger and S.~J. Brodsky,
\newblock Phys. Rev. Lett. {\bf 42}, 940 (1979).
%%CITATION = PRLTA,42,940;%%

\bibitem{Brodsky:1994kg}
S.~J. Brodsky, M.~Burkardt and I.~Schmidt,
\newblock Nucl. Phys. B {\bf 441}, 197 (1995).
%%CITATION = HEP-PH/9401328;%%

\bibitem{Gluck:1999xe}
M.~Gl{\"u}ck, E.~Reya and I.~Schienbein,
\newblock Eur. Phys. J. C {\bf 10}, 313 (1999).
%%CITATION = HEP-PH/9903288;%%

\end{thebibliography}

\end{document}